\documentclass[11pt]{amsart}
\usepackage[utf8]{inputenc} % If utf8 encoding
\usepackage[T1]{fontenc}    %
\usepackage[english]{babel} % English please
\usepackage{amssymb, mathtools} % Math

\usepackage{amsfonts}
\usepackage{amsmath}
\usepackage{color}
\usepackage{hyperref}

\usepackage{rotating}
\usepackage{braket}

\usepackage{stackrel}

\usepackage{pict2e}
\usepackage{graphicx}
\numberwithin{equation}{section}
\allowdisplaybreaks[4]

	\newcommand{\be}{\begin{eqnarray}}
\newcommand{\ee}{\end{eqnarray}}

\newcommand{\nn}{\nonumber}

\newcommand{\mat}{\left ( \begin{array}{cc}}
	\newcommand{\emat}{\end{array} \right )}
\newcommand{\vect}{\left ( \begin{array}{c}}
	\newcommand{\evect}{\end{array} \right )}

\newcommand{\e}{{\rm e}}
\newcommand{\dd}{{\rm d}}

\newcommand{\eins}{\leavevmode\hbox{\small1\kern-3.8pt\normalsize1}}

\begin{document}
	\title[Interactions Between Different Birds of Prey as a Random Point Process]
	{\bf{
			Interactions Between Different Birds of Prey\\ as a Random Point Process
	}}
	
	\author[G.~Akemann et al.]{Gernot Akemann$^1$, Nayden Chakarov$^2$, Oliver Kr\"{u}ger$^2$,\\Adam Mielke$^3$, Meinolf Ottensmann$^2$, and Patricia P\"{a}\lowercase{\ss{}}ler$^1$}
	\address{${}^1$Faculty of Physics and ${}^2$Faculty of Biology,  
	Bielefeld University,  P.O. Box 100131, D-33501 Bielefeld, Germany;
	\newline 
	${}^3$DTU Compute, Richard Petersens Plads, bygning 324,	2800 Kgs. Lyngby, Denmark} \email{akemann@physik.uni-bielefeld.de, oliver.krueger@uni-bielefeld.de,\newline  
	patricia@physik.uni-bielefeld.de, admi@dtu.dk}

\begin{abstract}

The two-dimensional Coulomb gas is a one-parameter family of random point processes, depending on the inverse temperature $\beta$. 
%%%%%%%
{
Based on previous work,} %successful application,} 
it is proposed as a simple statistical measure to quantify the intra- and interspecies repulsion among three different highly territorial birds of prey.
%, by comparing to the spacing between their nests in the plane. 
Using data from the area of the Teutoburger Wald over 20 years, we fit the nearest and next-to-nearest neighbour spacing distributions between the respective nests of Goshawk, Eagle Owl and the previously examined Common Buzzard to $\beta$ of the Coulomb gas. Within each species, the repulsion measured in this way deviates significantly from the Poisson process of independent points in the plane. 
In contrast, the repulsion amongst each of two species is found to be considerably lower and closer to Poisson. Methodologically we investigate the influence of the terrain, 
of a shorter interaction range given by the two-dimensional Yukawa interaction, and the statistical independence of the time moving average we use for the yearly ensembles of occupied nests. 
%%%%
{
We also check that an artificial random displacement of the original nest positions of the order of the mean level spacing quickly destroys the repulsion measured by $\beta> 0$.} 
A simple, approximate analytical expression  for the  nearest neighbour spacing distribution derived from non-Hermitian random matrix theory proves to be very useful.
%, being valid for the two-dimensional Coulomb gas close to Poisson at $\beta=0$.

\end{abstract}
 
%\date{\today}
\maketitle

%%%%%%%%%%%%%%%%%%%%%%%%%%%%%%
\section{Introduction} \label{Sec:Intro}

The study of the statistics of random point processes in one (1D) and two dimensions (2D) is a very active area of research, with many applications in physics and other sciences. Examples in 2D, on which we will focus here, include quantum optics \cite{QOreview}, quantum chaos \cite{Ueda,Pandey,AKMP,SaRibeiroProsen}, condensed matter physics \cite{TS}, statistics \cite{Moller} and ecology \cite{Clark,Law09,Buzzard}.
The goal of such an analysis is to clarify, first, whether the points coming from experimental data or numerical simulations are independent or not, and if not to quantify their correlations. The former case is called Poisson point process, where the points are distributed independently on a line in 1D, or in the plane in 2D. Closed form expressions exist for arbitrary dimension $D$, cf. \cite[Appendix A]{SaRibeiroProsen}.
Many point processes found in nature show correlations, and in particular repulsion between points in a characteristic, universal way, such that simple models from statistical mechanics apply. Before describing the point processes we use in more detail, let us explain what kind of understanding could be expected from such an approach to biological systems, which typically have a high degree of complexity.

A central question in biology is to understand the distribution of animals or plants in space and time. Furthermore, one would like to disentangle the effect of direct interaction within one species, between different species, and the effect of the environment. Such an understanding is of great value for conservation etc. 
Competition is one of the most ubiquitous features of ecology \cite{Begon}. Animals compete for limited resources with individuals of their own (intraspecific competition) or another species (interspecific competition), see \cite{Krueger2002a}. One of the easiest and hence most often used estimates of competition is the nearest neighbour distance \cite{Clark}. It has been shown to be a useful measure, with the underlying assumption that a too near neighbour of the same or another species ultimately leads to decreased reproduction or survival \cite{CK10,MCHK}. 

Two examples where 2D point processes have been applied in biology are the spacial distribution of trees, see \cite{Law09} for a review, and the distribution of nests of birds of prey in space and time in our previous work \cite{Buzzard}. Using data from \cite{KCNLGS-JM,CBK,MCHK} on the yearly locations of occupied nests of the Common Buzzard in an area of 300 km$^2$ in the Teutoburger  Wald, Germany, cf. Section \ref{Sec:data},  we have been able to draw the following conclusions. The distribution of their nests differs significantly from 2D Poisson statistics. We have been able to quantify the repulsion between the locations of nests of these highly aggressive birds of prey, by fitting the spacing distribution between nearest (NN) and next-to-nearest neighbours (NNN) in radial distance to those of a static 2D Coulomb gas of point charges in a confining (Gaussian) potential. Such a point process is also called Gibbs ensemble, cf. \cite{Moller}.
The single fit parameter used is the coupling strength $\beta$, representing the inverse temperature in the 2D Coulomb gas. Details about this  approach will be given in Section \ref{Sec:2DC}.
Moreover, with this simple parametrisation we have been able to identify a change in time of the interaction strength measured by $\beta$, over the observed period of 20 years, as a function of the increasing population density. 
Let us emphasise that the fitted value  of $\beta$ does not have a direct biological meaning, but merely serves to quantify the repulsion, in particular the deviation from 2D Poisson statistics at $\beta=0$. 

%%%%
{
Currently, we do not have an underlying microscopic model that would lead to a NN spacing distribution between the nests of birds of prey of (approximately) 2D Coulomb type. 
In 1D, a model for the spacing between parked cars and birds on a power line has been developed in \cite{Seba}, 
{
leading to a Gamma-distribution. 
It differs from Hermitian random matrix theory with a logarithmic Coulomb interaction in 1D, to be described below.
Ref. \cite{Seba}} 
is based on independently distributed symmetric spacings and provides a link to the perception of space by humans and birds. Because the absolute ordering into consecutive spacings between cars (birds) in 1D is absent in 2D, a simple generalisation of such an approach is not possible. Instead, our argument for using the spacing distribution of the 2D Coulomb gas is based on universality, as it has been observed to apply in several very different examples in physics:
in dissipative quantum spin chains \cite{Ueda,AKMP,SaRibeiroProsen}, the dissipative kicked rotor \cite{Ueda,Pandey}
and Quantum Chromodynamics with quark chemical potential \cite{Tilo}.
This is of course a strong assumption for the biological system, but it has worked surprisingly well for the Common Buzzard \cite{Buzzard}.
Other authors make different strong mathematical assumptions when modelling point processes from biology, for example that they are determinantal \cite{Moller}. The data are then fitted by a presumed correlation kernel, which relates to the two-point function.
A plethora of models have been formulated to explain intra- and interspecific competition and density dependence in ecological time series, and they commonly make a number of ecological assumptions. Here, we test how powerful a simple approach from statistical mechanics might be, that does not carry any ecological assumptions with it.}
%%%%%%%

A prime example from physics for 
%such 
a transition from Poisson to correlated random variables is the Bohigas--Giannoni--Schmit or quantum chaos conjecture \cite{QCC1a,QCC1b,QCC2}. It relates to a further example for point processes that is frequently used in physics \cite{GMW}, the eigenvalues of random matrices \cite{Mehta}. For self-adjoint matrices these 
eigenvalues 
are real (1D), whereas non-Hermitian matrices have  complex eigenvalues in 2D. In both cases the eigenvalues are strongly correlated random variables, originating from independently distributed matrix elements in the classical Gaussian ensembles.
The quantum chaos conjecture states that eigenvalue statistics of the Hamiltonian of generic integrable quantum systems will follow Poisson statistics, according to Berry--Tabor \cite{BT}, while that of fully chaotic Hamiltonians obeys random matrix statistics (distinguished by their global symmetry under time reversal). Here, the NN spacing distribution between eigenvalues has played an important role to quantify the path from integrability to chaos in 1D, and ample numerical 
%and analytical 
evidence has been collected, 
%%%%
{
cf. \cite{GMW} and references therein. It is based for example on 740 spacings in the Sinai billiard, and 
1726 spacings from nuclear scattering data. 
A complete analytical picture based on a semi-classical expansion has taken over 25 years to be developed, cf.
\cite{Haake} and references therein. 
The quantum chaos} 
%This 
conjecture  has been extended to 2D for dissipative quantum systems \cite{GHS1988}, and 
%%%%
{
so far numerical} 
%%%%%%%
evidence has been presented, e.g.  from the spectrum of the corresponding Liouville operator \cite{Ueda,AKMP,SaRibeiroProsen}.

The repulsion of random matrix eigenvalues is very well understood. It follows the logarithmic 2D Coulomb interaction, both for real eigenvalues in 1D and complex ones in 2D, with only particular values for $\beta=1,2,4$ occurring in 1D, and $\beta=2$ in 2D \footnote{This holds for the classical Ginibre ensembles, see a further discussion in \cite{Ueda,AMP}.}. The reason is that random matrix eigenvalues represent determinantal or Pfaffian point processes and are thus integrable, in the sense that all eigenvalue correlation functions are explicitly known \cite{Mehta,ForresterCoulomb}. Their 
{
mathematical} 
universality in the limit of a large number of points $N$  (matrix dimension) 
has been proved 
%%%%%
rigorously 
in 1D, see \cite{Arno} and references therein, and in 2D, cf. \cite{Hedenmalm} and  \cite{AKMP} for the NN spacing distribution. This means that they do not depend on the choice of a Gaussian distribution of matrix elements. 
The quantum chaos conjecture has raised the obvious question of how to interpolate between the Poisson point process, that can be viewed as a special case of diagonal random matrices at $\beta=0$,  and the specific $\beta$-values occurring for random matrices. In 1D, 
heuristic, approximate descriptions exist, e.g. the Brody distribution, see \cite{GMW} for other proposals. In 2D, it has been proposed \cite{AKMP} to directly use the 2D Coulomb gas at intermediate values of $\beta$, and to determine the NN and NNN spacing distributions numerically, that can then be compared with data. This is 
%ese findings are 
reviewed in Section \ref{Sec:2DC}, in particular a corresponding approximate surmise for the 2D NN spacing distribution valid for small values of $\beta$ \cite{AMP}, cf. Appendix \ref{Sec:surmise}.
These NN and NNN spacings were then used for fits as a function of $\beta$, for the spacing of eigenvalues of the Liouville operator of dissipative, boundary driven systems in 2D \cite{AKMP}, and in the yearly spacing distributions of nests of the Common Buzzard \cite{Buzzard}. 
Here, every year represents a realisation of the ensemble, and we have taken averages over all years, or windows of time moving averages over several consecutive years, in order to be sensitive to the time dependence.

The goal of this article is to 
%%%%%%%
{ extend  
%go beyond 
the analysis of a single species \cite{Buzzard} to two further species, and to quantify their respective interactions.}  
For the three species of competing birds of prey living in the observed area, the Common Buzzard,  Goshawk and Eagle Owl, annual  data for the locations of their nests over the same area and period of time have been collected 
\cite{CBK,MCHK,CPK,Mueller}. 
Our goal is to try to quantify the interspecies interaction between each two species, and to compare it to the intraspecies interaction found within each species individually. Therefore, in a first step we have repeated the analysis from \cite{Buzzard} for Goshawks and Eagle Owls individually, see Section \ref{Sec:indiv}. This includes the dependence on the respective change in population 
%%%%
{ 
over 20 years, ranging between 12-29 and 1-23 pairs, respectively.} 
%%%% 
Because the spacing distribution for Eagle Owls differs considerably, whether they nest in the Teutoburger Wald (F) or in the adjacent plains (P), we have further split them into 2 groups. Due to a comparably low statistics we have been unable to address the time dependence for the interspecific interaction between different species so far, to be discussed in Section \ref{Sub:interact}. 

%%%%%%
{
While this study might be perceived as rather limited with regard to the sample sizes and time spans involved, each species-specific time series with complete coverage of over 20 years in a study area of 300 km$^2$ is already amongst the largest data sets for any predatory bird \cite{KCNLGS-JM},  
and the inclusion of three species from the same area over such a time-span and detail is probably unique in the world. Over the course of the study, 3377 breeding attempts of Common Buzzard have been documented, over 423 Goshawk breeding attempts and 
174 Eagle Owl breeding attempts, cf. Appendix \ref{Sec:numbers}.}
%%%%%

A further purpose of the present work is to analyse methodological issues in Section \ref{Sec:Method},  that came up in discussions when presenting the results of \cite{Buzzard}. Because all 3 species discussed only nest in forested patches, a first question addressed in Section \ref{Sec:PoissonForest} is whether the repulsion we observe within a species via a fitted $\beta\neq0$ is due to the (fractal) area of the forest, or represents a true repulsion. We have generated a Poisson point process on the given forest area in the monitored region, and made a fit to a real $D>0$, for a $D$-dimensional Poisson NN spacing distribution. As we will see, this makes the repulsion between nests even more pronounced, and is clearly not an effect of the local area.
In \cite{Buzzard} the NN and NNN spacing distributions of nests were found to be described by different (typically decreasing) values of $\beta$, clearly indicating a shorter interaction range than 2D Coulomb. Therefore we have investigated numerically the spacing distribution of the Yukawa interaction in 2D in Section \ref{Sec:Yukawa}. It depends on 2 parameters, with a shorter interaction than logarithmic at larger distance. It reduces to 2D Coulomb in a limiting case and at small distance. 
Furthermore, the question of "independence" of nests occupied in consecutive years is analysed in Section \ref{Sec:Reuse}. Here, we quantify the reusage of nests within one or all species. A large abundance of old nests exists in the area, and the precise locations of nests in the monitoring process allows to quantify the respective percentage of reusage.

For a comparison to the NN or NNN spacing distribution from the 2D Coulomb gas, the mean density of the data has to be normalised to unity, a procedure called unfolding. For 1D this is unique and easier than in 2D, cf. \cite{GMW} vs. \cite{Tilo,AKMP}. As an alternative quantity that does not require unfolding, spacing ratios have been introduced in 1D \cite{Huse} and 2D \cite{SaRibeiroProsen}. However,  they are only known analytically for Poisson and for random matrix statistics, the complex Ginibre ensembles in 2D \cite{SaRibeiroProsen,DusaWettig}. In contrast to 1D \cite{Vivo2}, 
where the spacing ratio was derived as a function of $\beta$,
the interpolation in 2D is not so clear, as we will see from our data analysis presented in Appendix \ref{Sec:Ratios}. 
%%%%
{
In Appendix \ref{Sec:Displacement} we check that a random displacement of the observed positions of the nest by the order of the mean level spacing destroys the repulsion in the NN spacing and quickly leads to a Poisson distribution.} 
%%%%%%%5
All our findings, open questions and possible future directions are discussed in Section \ref{Sec:Conclusio}.

%%%%%%%

%%%%%%%%%%%%%%%%%%%%%%%%%%%%%%
\section{Description of Setup: Data and Point process} \label{Sec:Setup}

\subsection{Biological description and data collection - birds of prey in the Teutoburger Wald}\label{Sec:data}

\begin{center}
	\begin{figure}[t]
		\centering
		\begin{minipage}{0.47\linewidth}
			\includegraphics[width=\linewidth,angle=0]{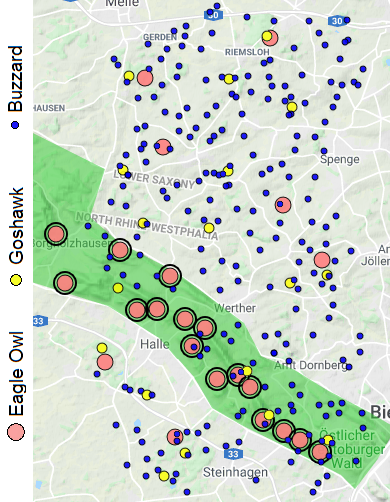}	
		\end{minipage}
		\hspace{10pt}
		\begin{minipage}{0.44\linewidth}
			\includegraphics[width=\linewidth,angle=0]{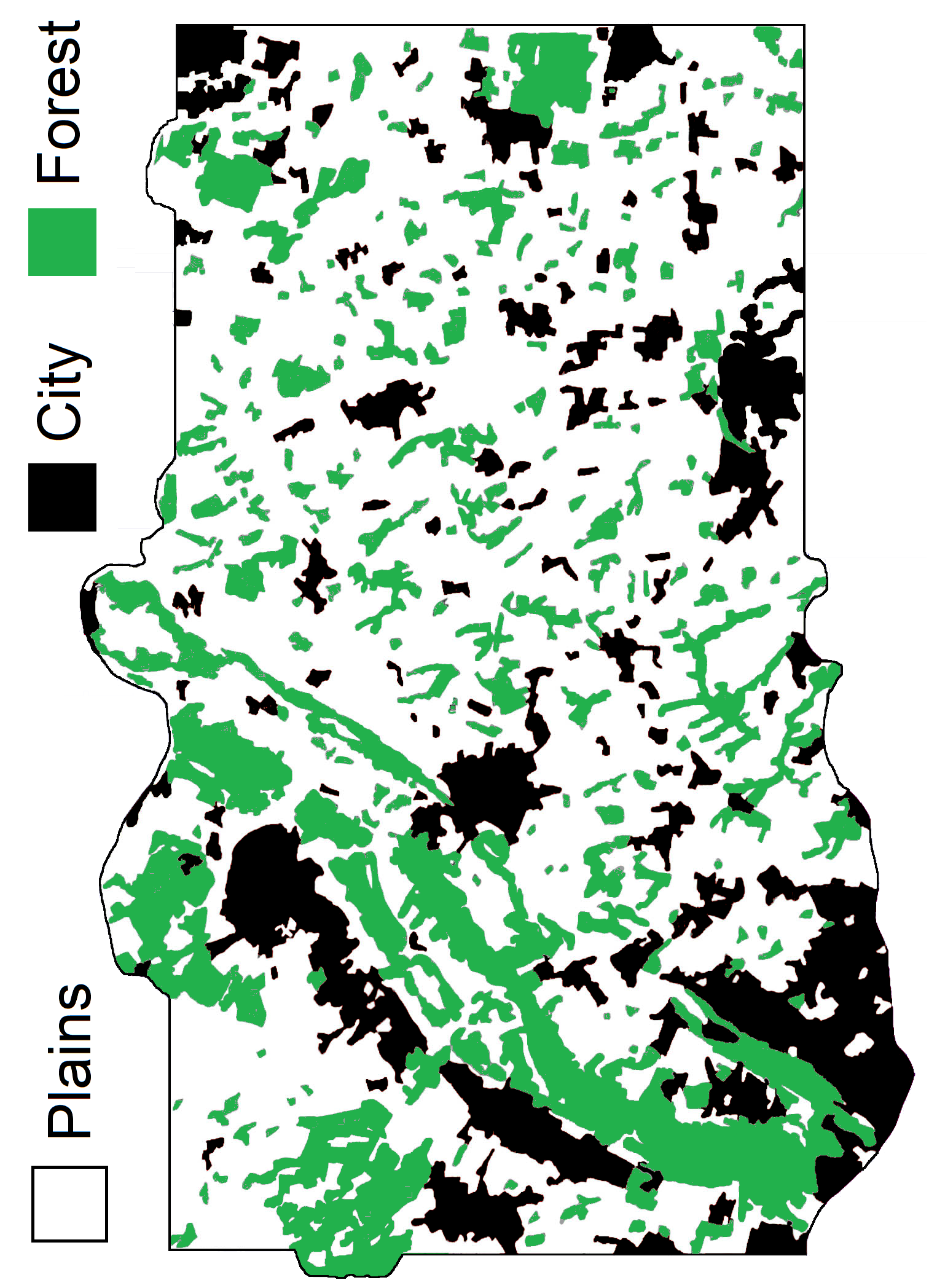}
		\end{minipage}
		\hspace{10pt}
		\caption{
\textbf{Left:} Distribution of all observed occupied bird nests in the year 2020: Eagle Owls (large pink points), Goshawks (yellow points) and Common Buzzard (dark blue points). There exist occupied nests outside the area with marked points, which have not been monitored. The green shaded area marks the approximate extend of the Teutoburger Wald based on the roads on either side, cf. right plot.  Eagle Owl nests within this green area are distinguished by a black ring around the large pink points.
{
 The extent of the monitored area is roughly $12\times25$ km.
}
\newline
		\textbf{Right:} A classification of the observed area in three types of terrain: City (black), forest (green) and cultivated land in the plains (white). 		
		The lack of population of birds within the smaller cities of Halle or Werther is well visible on the left plot. The forest found in the North of the Teutoburger Wald in that area consists of many small patches of irregular shape.		
		}
		\label{Fig:PlotBirds}
	\end{figure}
\end{center}

In this subsection we describe the collection of data we use and the geographical setup. The locations of occupied nests of three kinds of birds were collected in late winter and early spring during the years 2000-2020 in an area in and around the Teutoburger Wald close to Bielefeld. The three species are the Common buzzard (Buteo buteo L.), Goshawk (Accipiter gentilis) and Eagle Owl (Bubo bubo). 
The area of 300 km$^2$ (8$^\circ$25'E and
52$^\circ$6'N) is located in Eastern Westphalia. It consists of two
125 km$^2$ grid squares and 50 km$^2$ edge areas, see Fig. \ref{Fig:PlotBirds} for illustration. 
The main habitat of these birds of prey is
the Teutoburger Wald and a cultivated landscape to
the north and south of it. This is a low mountain region of height up to 315 m
above sea level, and we will treat the area as approximately flat, that is two-dimensional. 
All three species nest in forest patches. Their size  varies from rows of trees to large
patches, of  more than 10 km$^2$. In total approximately ~17\% of the study site is
forested. The area  has been
intensively monitored for birds of prey, and the resulting spatial
data have been published and used before extensively~\cite{KCNLGS-JM,CBK,MCHK}.

The forest patches were visited in March and April each year to look for incubating birds, and if a nest was occupied in both months, the pair was classified as breeding. The  locations of these nests were marked in large scale maps or using GPS-devices to monitor the spacial distribution of occupied nests. In Fig. \ref{Fig:PlotBirds} left, a snapshot for the year 2020 is shown, with occupied nests of the three species marked with coloured points, see caption. The classification of the monitored area into forest, city or plains is shown in  Fig. \ref{Fig:PlotBirds} right.

The example year 2020 chosen has amongst the highest population for all three kinds of birds. Already by eyesight several features emerge. Common Buzzards are much more abundant than Goshawks and Eagle Owls. Whereas Common Buzzards and Goshawks spread relatively evenly over the monitored area, the density of Eagle Owls inside  the Teutoburger Wald is much higher than outside.
Because of this difference in distribution of Eagle Owl nests, we decided to split the group of Eagle Owls in two: those nesting in the forest area (F) in green in the left  plot (nests marked by a black rings around the pink dot), and those in the plains (P) (nests marked by a pink dot only). Notice that the plains were populated by Eagle Owls only after 2010, see Fig. \ref{Fig:EagleOwlGroup} bottom right
%%%%%%
{ and Appendix \ref{Sec:numbers} Table \ref{tableNumbers}} 
 for their annual populations. The marking of the forest area in green in Fig. \ref{Fig:PlotBirds} left is of course approximate. In particular, one might argue whether the two outliers of Eagle Owls South of the Teutoburger Wald should be counted as (F) or (P), in view of the two large patches of forest in Fig. \ref{Fig:PlotBirds} right. Although living in large forest patches, they are separated from the main forest by cities and are thus counted as (P) here.
Such a choice does not change the quantitative picture, with the low statistics for Eagle Owls being difficult anyhow, 
%%%%%
{ cf. Table \ref{tableNumbers}.}

A second feature is visible by eye: There are holes in the population densities around cities (marked in black in Fig. \ref{Fig:PlotBirds} right). 
Although effects of the edge of a density are known in point processes, see the discussion at the end of Subsection \ref{Sec:2DC} below, we will disregards them here and treat all points as bulk points.

The third feature that is immediate from Fig. \ref{Fig:PlotBirds} left is that the mean spacing between nests within one species is very different for Goshawks, Common Buzzards, Eagle Owls (F) and Eagle Owls (P). It is approximately given by the inverse density (for a constant density). 
%%%%%%
{
The mean spacing in kilometers, e.g. $m_B=0.63$ km for the Common Buzzard and $m_G=2.63$ km for the Goshawk in 2020, is very different and  
encodes important biological information about the range of interaction within one (or different) species. 
Goshawks are among the most territorial of all European birds of prey and have large territory sizes of commonly between 5 and 50 km$^2$ and regularly make foraging trips of 5-10 km distance 
\cite{Kenward}. 
Interactions between neighbouring territories are rather aggressive and intruders into the territory are regularly killed \cite{Kenward}. As the mean nearest neighbour distance in our study site 
$m_G=2.63$ km is very low compared to other study sites due to our high population density 
\cite{KCNLGS-JM},  regular antagonistic interaction between the territorial pairs is exceedingly likely.}
%%%%
However, we are interested in comparing with (universal) features from spatial statistics in simple 2D point processes. Such a comparison can only be made quantitative if the mean density is normalised, and the fluctuations around this mean, as given for instance by the local spacing distribution, is measured and compared with the predictions from (equally normalised) point processes. 

The procedure of normalising the mean density is called unfolding and very well studied in applications of random matrix theory, see \cite{GMW}. For data in one spatial dimension it is unique and mostly straight forward, see \cite[Sect. 3.2.1]{GMW}. In two spatial dimensions it is more delicate, see \cite{Tilo,AKMP} for different procedures. Here, we will apply the method proposed in \cite{AKMP}, 
%%%%%%
{ which goes as follows.} 
Taking for example the Common Buzzards only, each blue point in  Fig. \ref{Fig:PlotBirds} left is replaced by a Gaussian distribution of a certain width, the sum of which defines the approximate mean density $\rho_{\rm ave}$ of points in 2D. The measured spacing around point $z=(x_0,y_0)$ is then normalised by rescaling with $\sqrt{\rho_{\rm ave}(x_0,y_0)}$, to achieve a spacing with respect to a mean density of unity. This process is done for each year for each of the 4 sets of birds, recalling that we have split the Eagle Owls into 2 sets,  yielding an ensembles for each set. This can then be compared to random point process to quantify the degree of repulsion, to be described in the next subsection.

%%%%%%%%%%%%%%%%%%%%%%
\subsection{Random point processes - from Poisson to Coulomb gas in 2D}\label{Sec:2DC}

In this subsection we introduce the random point processes and their corresponding nearest neighbour (NN) and next-to-nearest neighbour (NNN) spacing distribution that we will use for a comparison to data for the distribution of occupied nests of birds of prey. 
%%%%%%%
{ We will mainly discuss the spacing distributions in the following. However, for fits 
we will exclusively use the cumulative distribution function (CDF), which is independent of the choice of binning into histograms for the data.} 

We begin with the Poisson (Poi) random point process in two dimensions, cf. Section \ref{Sec:PoissonForest} for general dimension $D$. It consists of $N$ independent, uncorrelated points in the plane. The following normalised spacing distributions in radial distance between NN and NNN are known in the limit of a large number of points $N$:
\begin{eqnarray}
   p_{\mathrm{Poi}}^{\mathrm{(NN)}}(s) & =& 
    \; \frac{\pi}{2} s\ \e^{-\pi s^2/4}\quad  \sim s\ ,
    \label{PoiNN}\\
   p_{\mathrm{Poi}}^{\mathrm{(NNN)}}(s) & =&
   \; \frac{\pi^2}{8} s^3\ \e^{-\pi s^2/4} \, \sim s^3\ ,
   \label{PoiNNN}
\end{eqnarray}
see e.g. \cite{Haake,SaRibeiroProsen} for details of the derivation. The NN distribution has its first moment normalised to unity to set the scale (for the NNN spacing the first moment follows from this scale and is given by $3/2$). These spacing distributions can be obtained by distributing $N$ uncorrelated points on a disc of radius $R$. In the limit $N\to \infty$, after rescaling the mean spacing between points as $\bar{s}=R\sqrt{\pi/(4N)}$, eqs. \eqref{PoiNN} and \eqref{PoiNNN} are obtained in units of $\bar{s}$. Despite resulting from uncorrelated points, the 2D area measure in polar coordinates, $dxdy=sds\,d\theta$ for $z=x+iy=se^{i\theta}$, leads to a linear (cubic) repulsion in the  NN (NNN) spacing distribution. We will not consider higher order spacings distributions in the following.
As we will see below, the situation of a constant uniform density on a disc can also be obtained for a 2D Coulomb gas. Let us emphasise that the above spacing distributions quantify 
%%%%%%
{
(the absence of)} 
local correlations amongst points at distance $\sim 1/\sqrt{N}$, compared to global distances of the order of unity on the (unit) disc.

Let us move to the 2D static Coulomb gas (Cou). For a finite number $N$ of (charged) points it is given by the equilibrium distribution at inverse temperature $\beta=(k_BT)^{-1}$, subject to the logarithmic long-range interaction and a confining potential $V(z)$. The latter is chosen to be Gaussian here for simplicity, $V(z)=|z|^2$, 
\begin{eqnarray}\label{Eq:Coulomb}
  \mathcal{P}^{}_{\mathrm{Cou}, \beta} (z_1,\dots,z_N) &=& \frac{1}{\mathcal{Z}_{N,\beta}}
   \exp\left[\beta\sum_{{j,k=1;j<k}}^N
   \log |z_k - z_j|-\sum_{j=1}^{N} |z_j|^2\right],
\end{eqnarray}
where $\mathcal{Z}_{N,\beta}$ is the normalising partition function. The $N$ points are represented by complex coordinates $z_j\in\mathbb{C}$, $j=1,\ldots,N$, with the usual identification $\mathbb{C}\sim\mathbb{R}^2$.  
Compared to standard conventions, where $\beta$ multiplies the entire Hamiltonian including the confining potential, we have absorbed $\beta$ in front of the potential by rescaling the  coordinates $\beta V(z)\to V(z)$. The rescaling is made in order to allow us to take the limit $\beta\to0$ to reach Poisson statistics, while maintaining a confining potential. 
Furthermore, the charge is set to unity (or absorbed into $\beta$). 

Such static Coulomb gases have been much studied in the recent mathematical literature, and we refer to \cite{Sylvia} for a recent review. 
%%%%%%
{
Other classes of repulsion are possible such as a power law decay also called Riesz gas, cf. \cite{Lewin} for mathematical results and references therein. Due to the universality of the 2D Coulomb gas mentioned in the introduction we will stick to this class of interactions.}
After a further rescaling of the potential $V(z)\to NV(z)$, in the large-$N$ limit the mean density $\rho(z)$ of points condenses on the so-called droplet $S$. It is given by the Laplacian of the potential, $\rho(z)=\frac{2}{\beta}\partial_z\partial_{\bar{z}}V(z)$, Frostman's equilibrium measure. For general rotationally invariant potentials $V=V(|z|)$ the support $S$ is given by a disc of fixed radius, that can be rescaled to the unit disc. 

For the local correlations among points at distance of order $1/\sqrt{N}$ very little is known 
%%%%%%
{ analytically} 
for fixed $\beta>0$, apart from Poisson statistics at $\beta=0$ (that extends to $\beta\sim 1/N$, cf. \cite{Gaultier}) and the integrable case $\beta=2$, when the point process \eqref{Eq:Coulomb} becomes determinantal, cf. \cite{ForresterCoulomb}. 
Inspired by the Wigner surmise for the 1D Dyson gas, and its generalisation to general $\beta$ based on a $2\times2$ $\beta$-ensemble \cite{PierSatya}, a surmise (sur) for the NN spacing distribution was derived from complex normal $2\times2$ random matrices in \cite{AMP}\footnote{Notice a typo in the normalisation constant in \cite{AMP}: $\alpha^\beta$ there should be replaced by $\alpha^{1+\beta/2}$ as here.}
\begin{equation}
	\label{Eq:betasurmise}
	{p}^{\rm (NN)}_{{\rm sur},\beta}(s)=
	\frac{2\alpha^{1 + \beta/2}}{\Gamma[1+\beta/2]}
	\,s^{1+\beta} \exp[-\alpha s^2]\sim s^{1+\beta},
\end{equation}
where $\alpha={\Gamma[(3+\beta)/2]^2}/{\Gamma[1+\beta/2]^2}$. Its behaviour at small values $s\to 0$ is as expected heuristically from \eqref{Eq:Coulomb}, with one power from the radial measure and a power $\beta$ from the Vandermonde determinant. Unfortunately, it is well known \cite{GHS1988} from the integrable case at $\beta=2$ in 2D, cf. \eqref{GinibreSpacingNN} below, that here $N=2$ is not a good approximation to the large-$N$ limit. However, in the limit $\beta\to0$ eq. \eqref{Eq:betasurmise} exactly reproduces the NN spacing of the Poisson distribution \eqref{PoiNN}. This characteristic is opposite to the Wigner surmise in 1D, which becomes more accurate for increasing values of $\beta$, rather than for $\beta\to0$.

\begin{center}
	\begin{figure}[h]
		\centering
		\includegraphics[width=0.45\linewidth,angle=0]{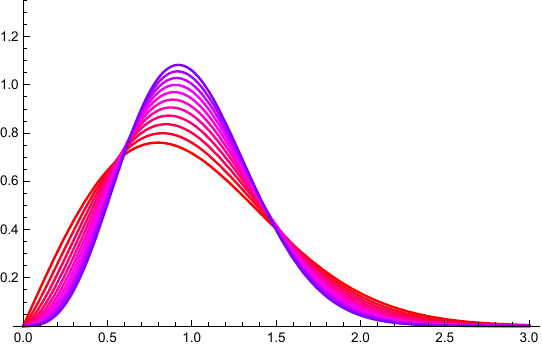}
				\includegraphics[width=0.45\linewidth,angle=0]{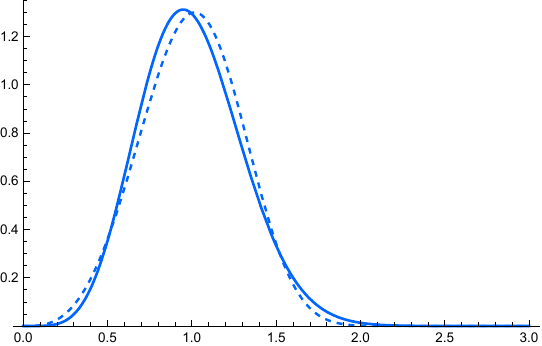}
		\caption{{\bf Left}: The approximate, surmised  NN spacing distribution ${p}^{\rm (NN)}_{{\rm sur},\beta_{\rm eff}}(s)$ from \eqref{Eq:betasurmise} and \eqref{Eq:betafit}, varying from $\beta=0$ (red) to $\beta=1$ (violett) in steps of 0.1. The maximum increases and moves from left to right.
		{\bf Right}: At $\beta=2$ the surmise 	(full line) is no longer close to the exact answer from the 		
		Ginibre ensemble \eqref{GinibreSpacingNN}, truncated at $N=20$ with rescaled, normalised first moment (dashed line).} 
		\label{Fig:Surmise}
	\end{figure}
\end{center}

In order to improve the approximation of the surmise \eqref{Eq:betasurmise} in 2D to larger values of $\beta$ (and with exact results for $N>2$ being unavailable), an effective $\beta_{\rm eff}$ was introduced in \cite{AMP}, by fitting a third order polynomial in $\beta$ to the numerically determined spacing of the 2D Coulomb gas in the range of $\beta\in[0,3]$:
\begin{equation}
\label{Eq:betafit}
\beta_{\rm eff}(\beta) = 2.108\beta - 0.190\beta^2 + 0.030\beta^3.
\end{equation}
Consequently in ${p}^{\rm (NN)}_{{\rm sur},\beta_{\rm eff}}(s)$ the normalisation constant has to change too, $\alpha(\beta)\to\alpha(\beta_{\rm eff})=\alpha_{\rm eff}$, to ensure a normalised spacing and first moment equal to unity. This leads a reasonable approximation up to $\beta\approx0.5$, with a standard deviation of up to $\sigma=2.6\cdot 10^{-2}$. See Appendix \ref{Sec:surmise} for plots in this range, and Fig. \ref{Fig:Surmise} right for a comparison at $\beta=2$, where the surmise clearly fails. In \cite{AMP} a more detailed comparison is presented, including higher $\beta$-values and their Kolmogorov-Smirnov distances. 
Obviously, the fit using an effective $\beta_{\rm eff}$ spoils the heuristically expected proportionality at very small argument $\sim s^{1+\beta_{\rm eff}}$. 
However, the limit $\beta\to0$ is still reproduced exactly.
Being based on $2\times2$ matrices, there is no approximate prediction for the NNN spacing possible.

In our comparison to data in the next section we will use the CDF of \eqref{Eq:betasurmise} 
%%%%%%
{ for the fitting procedure,} 
in order to avoid any dependence on the choice of histograms. It is given by 
\begin{eqnarray}
\label{Eq:CDFsurmise}
{E}^{\rm (NN)}_{{\rm sur},\beta_{\rm eff}}(s) &=& 1 - \frac{\Gamma\left(1 + \frac{\beta_{\rm eff}}{2}, \alpha_{\rm eff} s^2\right)}{\Gamma\left(1 + \frac{\beta_{\rm eff}}{2}\right)},
\end{eqnarray}
where $\Gamma(n+1, x)=\int_x^\infty t^n e^{-t}dt$ is the upper incomplete Gamma function. 

For completeness we give the analytical result for the NN \cite{GHS1988} and NNN \cite{APS} spacing distribution for $\beta=2$, that follows from the Ginibre ensemble \cite{Ginibre}. Here, the 2D Coulomb gas \eqref{Eq:Coulomb} has a representation in terms of complex eigenvalues of complex non-Hermitian random matrices with Gaussian distribution. In this case, the logarithmic interaction term can be written in terms of the modulus square of the Vandermonde determinant, 
\begin{equation}
\Delta_N(z_1,\ldots,z_N)=\det[z_i^{j-1}]_{i,j=1}^N=\prod_{k>l}^N(z_k-z_l)\ ,
\end{equation}
of the $N$ eigenvalues $z_j$. Hence the point process is determinantal, and all complex eigenvalue correlation functions are explicitly known at finite $N$. The spacing distributions can be derived from the limiting CDF or gap probability $E_{{\rm Gin}}(s)$, to find an eigenvalue at the origin and the closest non-zero complex eigenvalue at radial distance $s$,
\begin{equation}
E_{{\rm Gin}}^{\rm(NN)}(s)= 	\prod_{j=1}^{\infty}\frac{\Gamma(1 + j, s^2)}{j!}\ .
\end{equation}
For finite $N$ the product extends only to $N-1$. 
The  limiting spacing distributions at infinite matrix
dimension follow from differentiation, compare \cite{GHS1988} for NN and \cite{APS} for
NNN:
\begin{eqnarray}\label{GinibreSpacingNN}
	p^{\mathrm{(NN)}}_{\mathrm{Gin}}(s) &=&
E_{{\rm Gin}}^{\rm(NN)}(s)
	\sum_{j=1}^{\infty} \frac{2s^{2j+1}\e^{-s^2}}{\Gamma(1 + j, s^2)}
	\sim s^3, 
	\\
p^{\mathrm{(NNN)}}_{\mathrm{Gin}}(s) &=&
E_{{\rm Gin}}^{\rm(NN)}(s)
        \sum_{j,k=1; k \neq j}^{\infty}
        \frac{\gamma(1+j, s^2)}{\Gamma(1 + j, s^2)}
	\frac{2s^{2k+1}\e^{-s^2}}{\Gamma(1 + k, s^2)}	\sim s^5, 
	\label{GinibreSpacingNNN}
\end{eqnarray}
where 
$\gamma(1 + k, s^2)=\int^{s^2}_0 t^k \e^{-t}\dd t$ is the lower incomplete
Gamma function, and we give again the behaviour at $s\to0$, too. The products converge very rapidly and both spacing distributions are normalised to unity. For simplicity, we have given the expressions above where the first moment $\bar{s}_1$ of the NN spacing is not yet normalised to unity. It can only be determined numerically, for a product truncated at sufficiently high value of, say $N\approx 20$, or larger. The NN with first moment of unity is then obtained by rescaling $\hat{p}^{\mathrm{(NN)}}_{\mathrm{Gin}}(y)=\bar{s}_1 p^{\mathrm{(NN)}}_{\mathrm{Gin}}(\bar{s}_1y)  $, and the correspondingly rescaled NNN spacing reads $\hat{p}^{\mathrm{(NNN)}}_{\mathrm{Gin}}(y)=\bar{s}_1 p^{\mathrm{(NNN)}}_{\mathrm{Gin}}(\bar{s}_1y)  $.

The spacing distributions \eqref{GinibreSpacingNN} and \eqref{GinibreSpacingNNN} hold throughout the bulk of the spectrum inside the supporting unit disc and are known to be highly universal. That is they hold for a much larger class of confining potentials 
$V(|z|)$ than Gaussian, in general random normal matrix ensembles \cite{Hedenmalm}. 

In the Ginibre ensemble at $\beta=2$, the complex eigenvalue correlations at the edge of the droplet have also been investigated. They are also universal and agree with the correlations at an inner edge, that is a droplet with a hole, that can be studied in the induced Ginibre ensemble \cite{Fischmann}. In our data there is no outer edge in Fig. \ref{Fig:PlotBirds}, it is simply given by the area of observation, and there are birds nesting outside the area containing dots. However, we do observe inner edges, as the cities of Bielefeld, Halle or Werther show up as holes in Fig. \ref{Fig:PlotBirds}. We have not been able to address such edge correlations, mainly due to lack of statistics, but also because it is not so clear where to draw the boundaries of the inner edges. In our analysis we have thus treated all points 
%%%%%%
{ (nests)}  
as bulk points.

In the next section we will use both to the approximate NN spacing ${p}^{\rm (NN)}_{{\rm sur},\beta_{\rm eff}}(s)$ from the surmise \cite{AMP}, and the numerically determined NN and NNN spacing from \cite{AKMP} as functions of $\beta$, 
{
in steps of $0.1$, and we refer to \cite{AKMP} for details about the numerics. In particular around 200000 spacings have been used there to generate these numerical distributions.}

%in order to quantify the repulsion within one species, and between different species.  

%%%%%%%%%%%%%%%%%%%%%%%%%%%%%%
\section{Interaction of species} \label{Sec:Interact}

\subsection{Interaction within one species: Common Buzzard, Goshawk and Eagle Owl}
\label{Sec:indiv}

\begin{center}
	\begin{figure}[h]
		\centering
		\begin{minipage}{0.02\linewidth}
			\begin{turn}{90}\hspace{-20pt} \textbf{Eagle Owls (F)} \hspace{15pt} \textbf{Goshawks} \hspace{40pt} \textbf{Buzzards} \hspace{30pt}\end{turn}
		\end{minipage}
		\begin{minipage}{0.92\linewidth}
			\centering
			 \hspace{0.05\linewidth}\textbf{2020} \hspace{0.2\linewidth} \textbf{2011-2020} \hspace{0.15\linewidth} \textbf{2000-2020}\\
			\includegraphics[width=0.3\linewidth,angle=0]{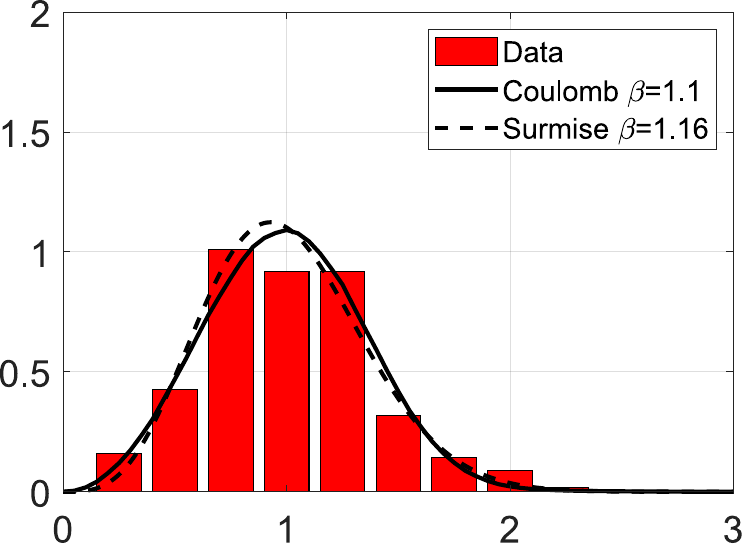}
			\includegraphics[width=0.3\linewidth,angle=0]{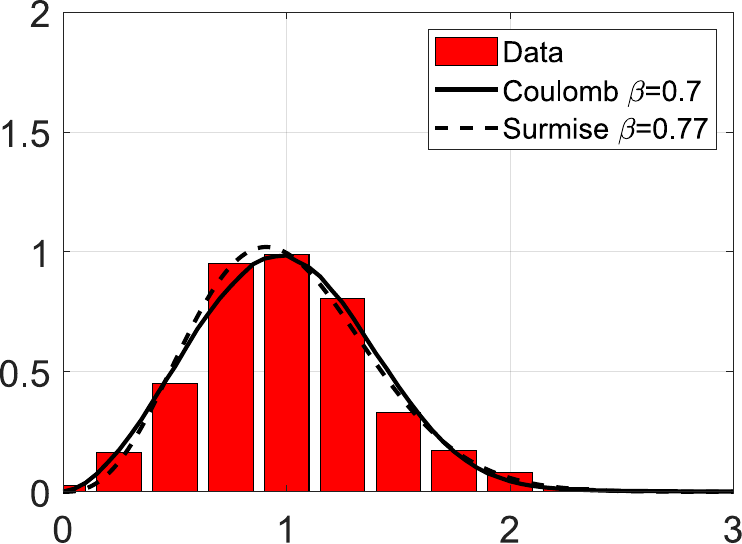}
			\includegraphics[width=0.3\linewidth,angle=0]{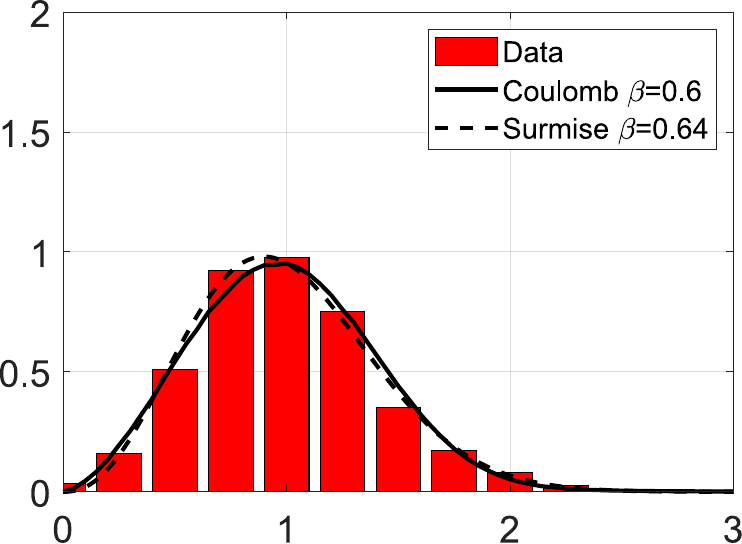}
			\\
			\includegraphics[width=0.3\linewidth,angle=0]{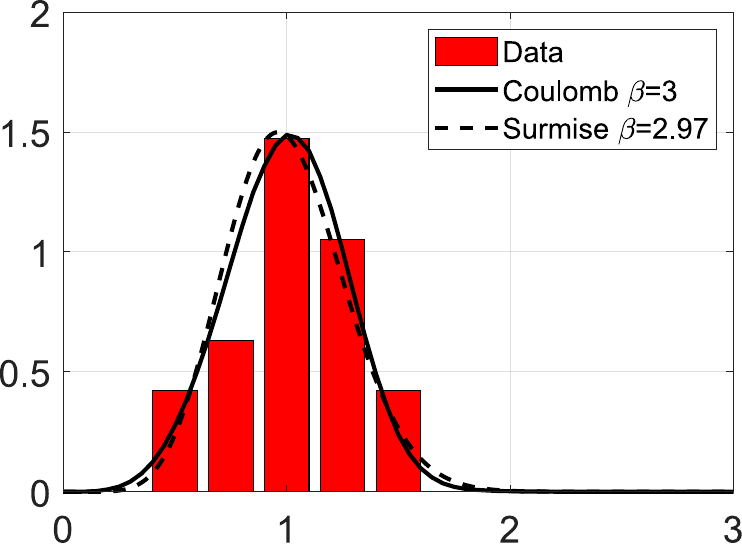}
			\includegraphics[width=0.3\linewidth,angle=0]{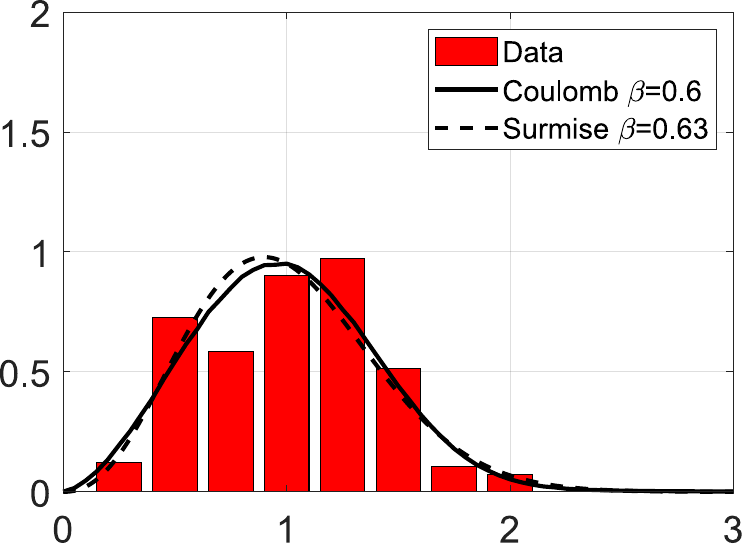}
			\includegraphics[width=0.3\linewidth,angle=0]{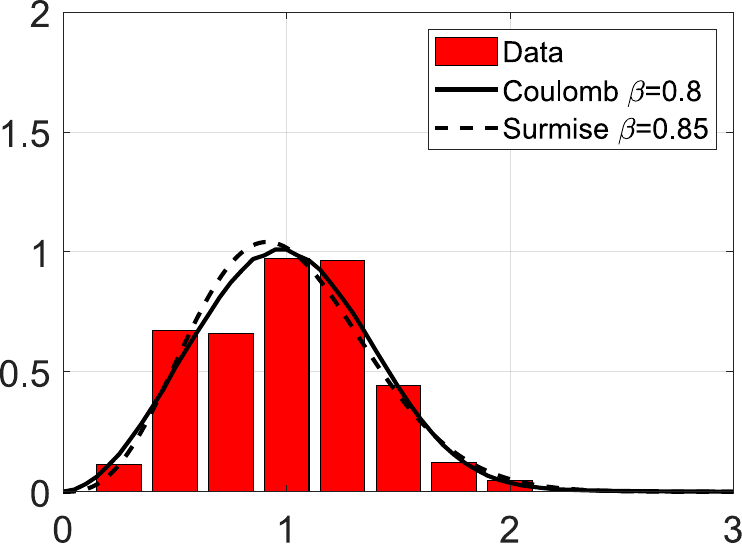}
			\\			
			\includegraphics[width=0.3\linewidth,angle=0]{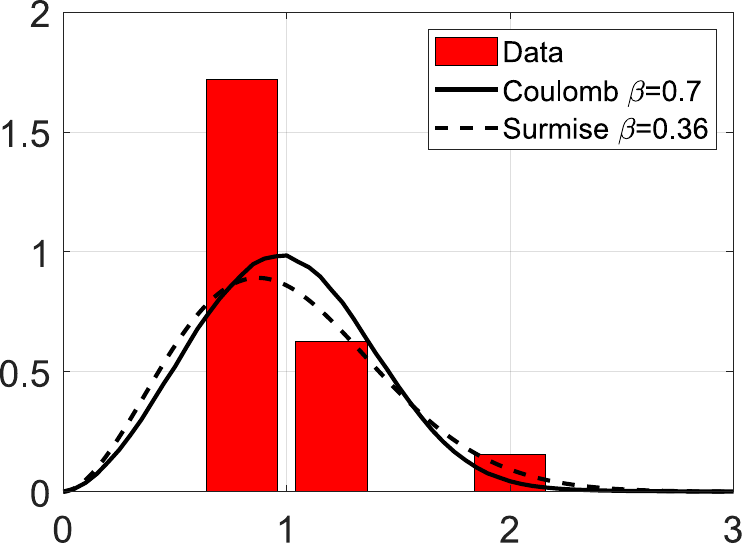}
			\includegraphics[width=0.3\linewidth,angle=0]{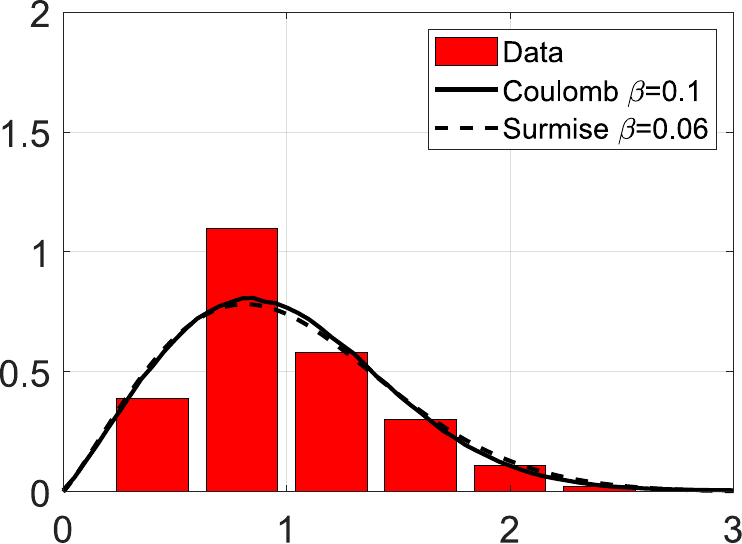}
			\includegraphics[width=0.3\linewidth,angle=0]{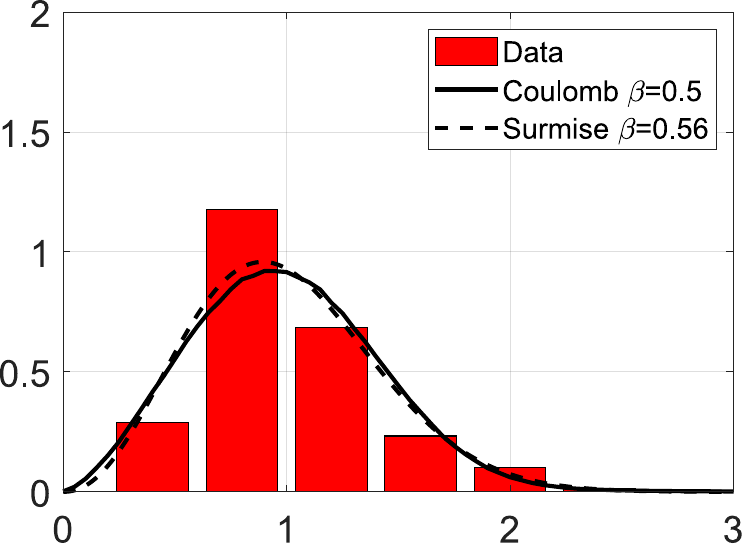}\\
		\end{minipage}
		\caption{Comparison of the fitted $\beta$-values from the NN spacing distribution from 2D Coulomb in steps of $0.1$ (full line), respectively from the surmise \eqref{Eq:betasurmise} and \eqref{Eq:betafit} with continuous $\beta$ (dashed line).	
	We present examples for fits for a single year 
%%%%%%%%%%%%%%%%%%
		2020 (left column), an average over 10 consecutive years from 2011-2020  (middle column), and over all years (right column). For the Common Buzzards (top row) we have 228 spacings in 2020, 2187 spacings in 2011-2020, and 3377 spacings in all years. The corresponding numbers of spacings 
for Goshawks (middle row) are 19, 226 and 423 respectively,  and for Eagle Owls (F,  bottom row) they read 16, 116 and 174,  cf. Table \ref{tableNumbers} in Appendix \ref{Sec:numbers} for details. 
	Shown are histograms and spacing distribution, but we emphasise that the fits were obtained using the cumulative distributions.
Clearly when moving from individual years to time moving averages, the number of spacings for Goshawks and Eagle Owls (F) leads to an improved statistics and quality of the fits.}
		\label{Fig:GroupingYears}
	\end{figure}
\end{center}

Let us describe the procedure initiated in  \cite{Buzzard} for Common Buzzards, to quantify the supposedly repulsive interaction between these birds of prey, by fitting the NN and NNN spacing distributions between occupied nests. Each year is observed for each species and treated as an ensemble. In principle, for each year a value of $\beta$ can be determined for each species, as it is done in Fig. \ref{Fig:GroupingYears} left column for the NN spacing distribution in the year 2020 as an example.  As it can be seen, the quality of the fit decreases rapidly with the amount of data available, in particular for the Eagle Owls. 
%%%%%%%
{
This has motivated us to consider time moving averages in the other columns, to be discussed below. 
In all figures}
we show both the fit using the Coulomb gas (full line), 
with the spacings determined numerically in \cite{AKMP} in discrete steps of size $0.1$ for $\beta$,
and the explicit formula \eqref{Eq:betasurmise} for the surmise (dashed line), with $\beta$ varying continuously.  
The fits are done using the CDF, compare \eqref{Eq:CDFsurmise}, to avoid any dependence on the choice of binning. To guide the eye we nevertheless show the spacing distribution compared to histograms of the data.
The fitted $\beta$-values vary over a wide range. For example, in the year 2020 in Fig. \ref{Fig:GroupingYears} left column,  we find the strongest repulsion among Goshawks ($\beta=3$), followed by the Common Buzzard ($\beta=1.1$) down to Eagle Owls (F) with $\beta=0.7$. Apart from the latter plot with very low statistics, the surmise and Coulomb gas fit agree well. 
Notice that the continuous
respectively discrete fit in $\beta$ adds to the discrepancy between the two, see Appendix \ref{Sec:surmise} for a comparison at equal values.
We find that all fitted values are far from 2D Poisson statistics at $\beta=0$. Let us emphasise, that the fitted parameter $\beta$ does not have a biological meaning, but that it is the relative strength that allows us to compare between different species of birds, or their interaction in the next Subsection \ref{Sub:interact}.
%%%%%%%
{
Without the successful application to the Common Buzzard in \cite{Buzzard} we would not have embarked onto an analysis of data from other birds of prey, with a considerably lower statistics. 
The inference of biological information from blood samples of the Common Buzzard with low statistics has been analysed in \cite{Mueller}.
}

\begin{center}
	\begin{figure}[h]
		\centering
		\hspace{0.05\linewidth} \textbf{$\beta$-fit Buzzards} \hspace{0.2\linewidth}
		\textbf{$\beta$-fit Goshawks}\\
		\includegraphics[width=0.4\linewidth,angle=0]{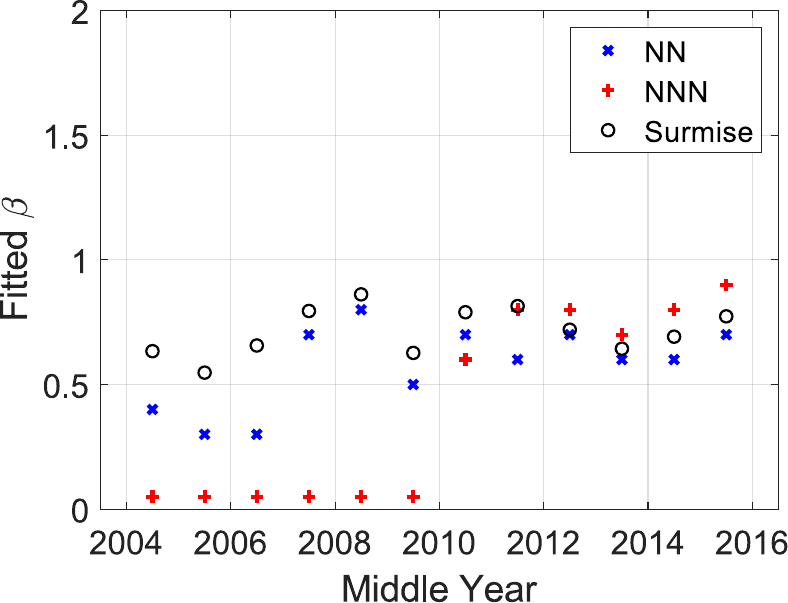}
\hspace{10pt}
		\includegraphics[width=0.4\linewidth,angle=0]{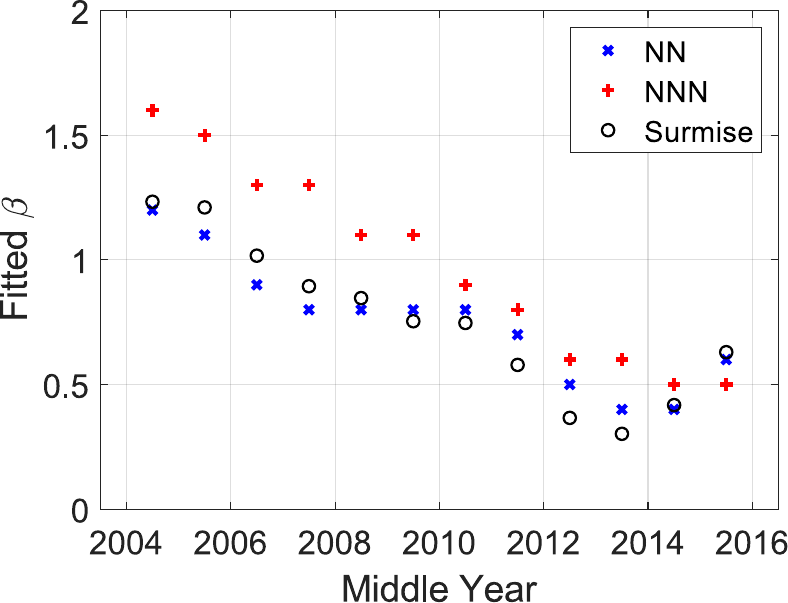}
\\
		\ \\
		\hspace{0.05\linewidth} \textbf{Common Buzzard population} \hspace{0.2\linewidth}
		\textbf{Goshawk population}\\
		\includegraphics[width=0.4\linewidth,angle=0]{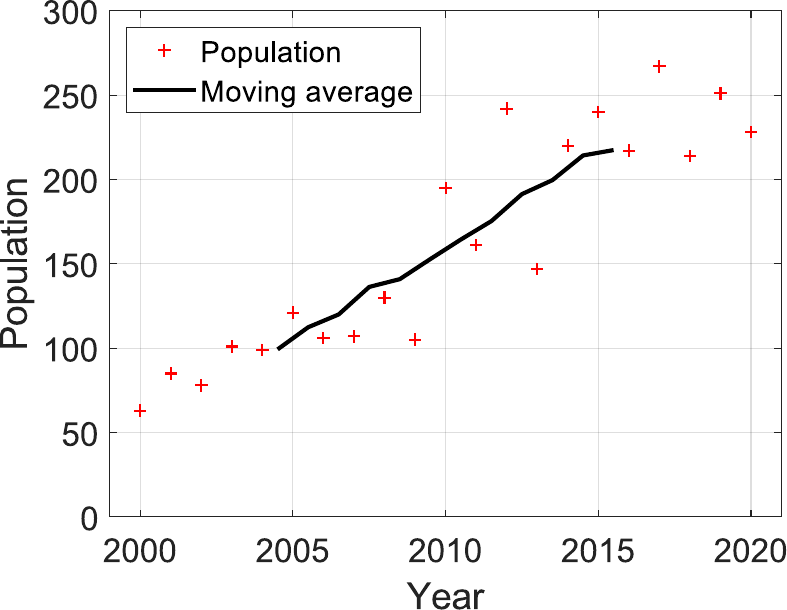}\hspace{10pt}
		\includegraphics[width=0.4\linewidth,angle=0]{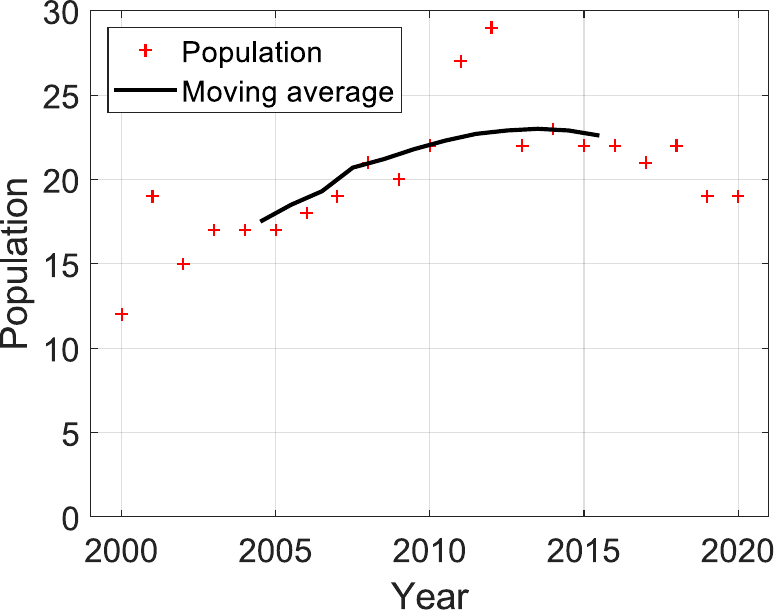}
		\caption{\textbf{Top row:} Fit of $\beta$ for the NN spacing for Common Buzzards (left) and Goshawks (right) from the Coulomb gas (blue crosses) and surmise (circles), as well as for the  NNN spacings distribution (red crosses). We use a time moving average of 10 years, where the middle year is indicated on the x-axis (e.g. 2004+1/2 for the period 2000-2009).
The number of spacings per fit ranges between 1050 %(2009)
and 2420 
%(2012) 
for the Buzzards, and between 170 and 290 
%(2012) 
for the Goshawks, as can be seen from the plots in the 		
		\textbf{Bottom row:} 
Population of birds per year (red crosses) for Common Buzzards (left) and Goshawks (right), over the entire period of time 2000-2020. The resulting average populations using a time moving average of 10 years is shown in comparison (black line), where again the middle year is chosen as a label, as in the top plots. For a better comparison these points are connected by the line of the moving average.}
		\label{Fig:AllBirds}
	\end{figure}
\end{center}

\begin{center}
	\begin{figure}[t]
		\hspace{0.01\linewidth} \textbf{$\beta$-fit Eagle Owls (F)} \hspace{0.2\linewidth}
		\textbf{$\beta$-fit Eagle Owls (P)}\\
\includegraphics[width=0.4\linewidth,angle=0]
{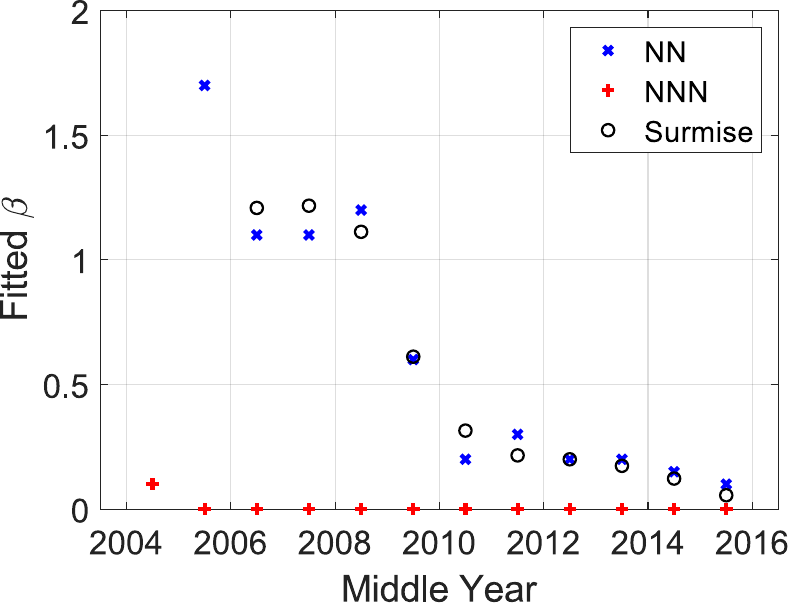}
		\hspace{10pt}
		\includegraphics[width=0.4\linewidth,angle=0]{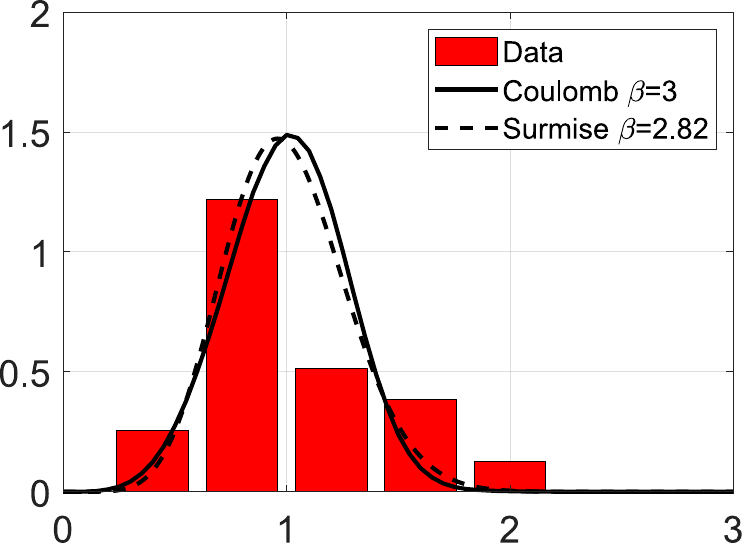}%
\\
		\ \\
			\hspace{0.01\linewidth} \textbf{Eagle Owl (F) population} \hspace{0.1\linewidth}
		\textbf{Eagle Owl (P) population}\\
		\includegraphics[width=0.4\linewidth,angle=0]{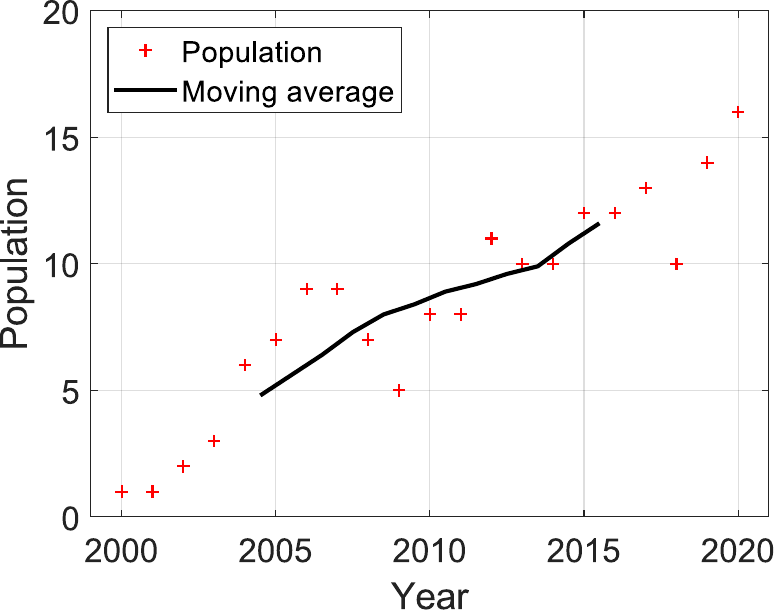}
		\hspace{10pt}
		\includegraphics[width=0.4\linewidth,angle=0]{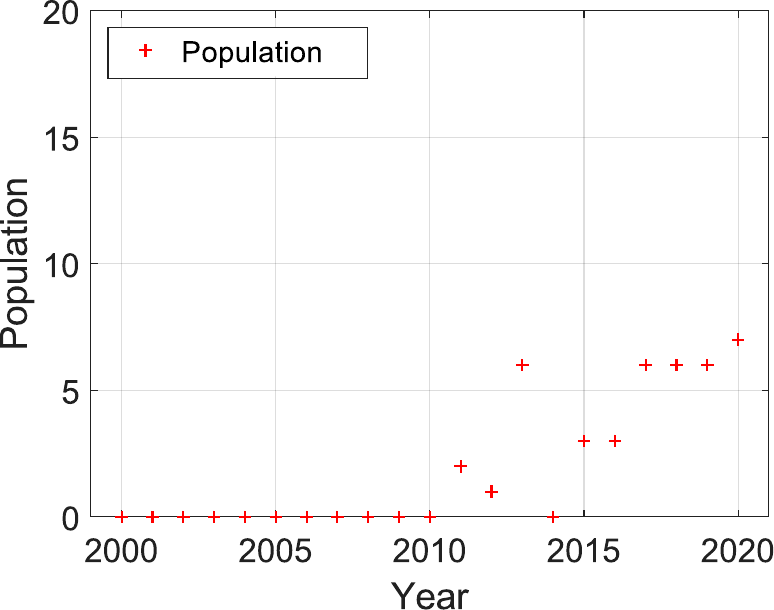}
				\caption{ The same plots as in Fig. \ref{Fig:AllBirds} for the Eagle Owls (F) are presented in the left column,  
with the corresponding $\beta$-fits for time moving averages (top left), and 
corresponding population with time moving averages in black (bottom left). Values above $\beta=2$ are not shown, and we have between 50 and 120 spacings for each fit.
Due to the lack of data for the Eagle Owls (P), see the corresponding population on the bottom right, we only present a fit over all years containing 40 spacings (top right), cf. Table \ref{tableNumbers}.  Notice the lack of nests in the plains (right) until 2010.
		}
		\label{Fig:EagleOwlGroup}
	\end{figure}
\end{center}

The population of pairs of Common Buzzards and Goshawks per year is shown as red crosses in Fig. \ref{Fig:AllBirds} bottom left, respectively right, and for Eagle Owls in Fig. \ref{Fig:EagleOwlGroup} bottom in the forest (F, left) and plains (P, right), 
%%%%%%%
{
cf. Table \ref{tableNumbers} in Appendix \ref{Sec:numbers}. 
Between Common Buzzards and Goshawks there is a factor of 10 in abundance, and 
typically} 
another factor of 2 between Goshawks and Eagle Owls, with the latter being completely absent in the plains up to the year 2010. 
To remedy the resulting low statistics, we have introduced a time moving average, averaging over 10 consecutive ensembles (years)  for all three species. The correspondingly averaged population is shown as a black line in the population plots in Figs. \ref{Fig:AllBirds} and \ref{Fig:EagleOwlGroup}, respectively. For comparison, 
because of the better statistics for the Common Buzzards, in  \cite{Buzzard} we choose a time moving average of 5 years, that gave a better resolution of the time dependence of the fitted $\beta$ as a measure of repulsion. 
In \cite{Buzzard} we also presented the fits for all individual years (compare Fig. 4 left in \cite{Buzzard}). This result is very noisy and makes it difficult to see an overall trend in time, which is why we do not present such plots here.

The fitted $\beta$-value for such time averaged NN spacing distributions for each species is shown in Fig. \ref{Fig:GroupingYears} middle column, choosing the average over the period 2011-2020 as an example, that includes the single year 2020 from the left column. It is striking to see that in all three cases the ensemble average leads to a much lower $\beta$-value. This is  especially true for the Goshawks, which now show a repulsion comparable to the Common Buzzards. 
There is a clear trend that the fitted $\beta$ value goes down for the Goshawks over the years, see Fig. \ref{Fig:AllBirds}. Notice that their population is peaked in 
%2011 and 
2012. 
For the Eagle Owls (F) we obtain a fitted value $\beta=0.1$, being very close to 2D Poisson in this window of time average. Also here the averaged fitted $\beta$ goes down over time, see Fig. \ref{Fig:EagleOwlGroup} top left.

Finally, if we abandon any time resolution and make an ensemble average over all available years, we obtain closer  values for $\beta$ for all three species, as shown in Fig. \ref{Fig:GroupingYears} right column. The strongest repulsion is again amongst Goshawks ($\beta=0.8$), followed by Common Buzzards ($\beta=0.6$), and almost equally among Eagle Owls (F) with $\beta=0.5$. 
The particularly high $\beta$ for intraspecific repulsion in Goshawk makes a lot of ecological sense, as the Goshawk is one of the most territorial species in Europe \cite{KL01}.  The Goshawk preys on larger prey and needs exclusive access to hunting areas \cite{KS96}. Therefore, intruding individuals are always met with high aggression and are regularly killed by the territory owners \cite{CS80,KL01,K07}, 
%%%%%%
{
cf. Fig. \ref{Fig:PlotBirds} for their relatively large spacings.} 
 All values are significantly above 2D Poisson statistics ($\beta=0$).
Because of the strong growth in population for Common Buzzards, and for Eagle Owls (F), see Figs.  \ref{Fig:AllBirds} and \ref{Fig:EagleOwlGroup}, respectively,
%%%%%%%
{
and Table \ref{tableNumbers} Appendix \ref{Sec:numbers},} 
the later years are weighted much higher here in the average over all years, because they contribute to more spacings per year.

In the following  we will discuss our findings for the $\beta$-values fitted to the time moving average. Two general observations can be made. First, the values obtained using the surmise agree quite well with the NN spacings form the 2D Coulomb gas, apart from the years with low statistics for the Eagle Owls (F). 
In particular, they follow the same trend as seen from the NN Coulomb gas fits and thus can be used as a very simple approximate, but analytical tool. Second, the $\beta$-values obtained from NN and NNN fits do not agree in general, sometimes lying systematically below (Common Buzzards, Eagle Owls (F)) or above (Goshawks) the NN 
value. This discrepancy may be used to compare the observed interaction range to that of the Coulomb interaction. 
The same  observation was already made in \cite{Buzzard} and motivated us to introduce and study an interaction with shorter range, the 2D Yukawa interaction, see Section \ref{Sec:Yukawa}. It turned out, however, that also with such a two-parameter family (with $\beta$ and $\gamma$) the phenomena of obtaining two different values for $\beta$ from NN and NNN remains, see the discussion there. Therefore,  we kept the simpler 2D Coulomb 
%%%%%%
{ interaction}
%gas model 
as a measure for repulsion.

In Figs. \ref{Fig:AllBirds} and \ref{Fig:EagleOwlGroup} we do not give error bars for the fitted values of $\beta$. In our previous paper \cite{Buzzard} we made an attempt to estimate the error by fitting $\beta$ to approximately the same number of points from a 2D Coulomb gas simulation, that is $N=200$ for the Common Buzzards. This estimate (which is not related to the data of occupied nests) seemed to overestimate the fluctuations, e.g. when comparing to a linear fit for the trend in $\beta$. Because in the ensembles of Goshawks and Eagle Owls (F) the number of nests per year is smaller by a factor of 10, we refrain from giving such rough estimates for the errors.  

For the Common Buzzards the following picture emerges from Fig. \ref{Fig:AllBirds} left, that was already found in \cite{Buzzard}. The NN values vary in range between $\beta=0.3$ and 0.8 over the observed period. Although they still fluctuate considerably for this time moving average,   an increasing trend can be identified. In contrast, the NNN values remain close to $\beta=0$ for half of the observed period, and then jump approximately to the NN values from the mid-year 2010 onwards. In comparison, the population growth is approximately linear from around 100 pairs on average to above 200 (10 year average given by the black curve in Fig. \ref{Fig:AllBirds} bottom left). Apparently below a certain population threshold the interaction range is rather limited to NN, and then increases in range to be more 2D Coulomb like up to NNN. 
Notice that because of unfolding the data, the mean spacing is the same in all years. Thus the trivial effect of having a smaller spacing for a larger density is removed here.

For the Goshawks we observe in Fig. \ref{Fig:AllBirds} right that there is a clear trend for the fitted $\beta$-value to go down from approximately 1.2 to 0.4 for NN and from 1.6. to 0.5 for NNN. In contrast to the Common Buzzards, the NNN values are systematically above the NN values, apart from the last middle year. This means that the interaction range is longer than 2D Coulomb, 
%%%%%%%
{ which goes in line with the high aggressiveness of this species discussed earlier.} 
The repulsion decreases over time, and the population development is also different from the Common Buzzards when comparing the bottom rows in Fig. \ref{Fig:AllBirds}. Notice the difference in abundance of a factor of 5-10 between the two species.
Although the Goshawk population fluctuates between 12 and 29 pairs, the average only slightly goes up from 17.5 to 22.5, stabilises and then slightly goes down again. Thus a different factor than the Goshawk population seems to be at work here. If it lies in the interaction between the two other species, it will be answered (negatively) in the next Subsection \ref{Sub:interact}.

Finally let us discuss the findings for the Eagle Owls. Because of the lowest statistics, see Fig. \ref{Fig:EagleOwlGroup} and Table \ref{tableNumbers}, about a factor of 2-10 in abundance below the population of Goshawks, this is very difficult, especially for the much smaller group nesting in the plains (P). 
For those nesting in the forest (F), the $\beta$ value from NN clearly decreases from large values above 1.5 down to 0.2-0.1, which is almost comparable with 2D Poisson at $\beta=0$. 
In contrast, the NNN values remain consistent with $\beta=0$ over the entire period. A comparison with the population development from 5 pairs on average to 12 shows an approximate linear increase on average, see Fig. \ref{Fig:EagleOwlGroup} bottom left. The effect is thus opposite to the Common Buzzards, the Eagle Owls (F) 
%(and Goshawks) 
show a decreasing repulsion, with short range interaction, despite an increase or stabilisation in population, respectively. 
%%%%%%%
{
It should be noted that for relatively few spacings of the order 20, 
%very few spacings of the order of unity, 
the distribution is strongly peaked, often leading to larger values for the fitted $\beta$. 
}
%%%%%%%%%%%%%%
For a better resolution the first NN value at $\beta=3$ is suppressed in
Fig. \ref{Fig:EagleOwlGroup} top left.

The time dimension of the strength of the repulsion also reflects the species' biology as well as population trends in the study area. In Common Buzzard, we see an increase of $\beta$ over time, most likely due to the increase in the population density until carrying capacity might have been approached in the last five years. The strength of intraspecific competition is expected to increase with increasing population density. With regard to the Goshawk, a decreasing beta over time is mirrored by a decrease in population density over the last five years, most likely due to displacement by Eagle Owls \cite{MCHK}. Why $\beta$ shows a decrease for Eagle Owls, the population of which has increased very rapidly over the last 20 years, is difficult to explain. It could be that the population is still not approaching carrying capacity and hence progressively smaller repulsions measured through NN spacings are observed.

One remark regarding 2D Poisson is in order here. In Subsection \ref{Sec:PoissonForest} we ask the question, if the scattered patches of forest (see Fig. \ref{Fig:PlotBirds}) where all birds prefer to nest, does not introduce an effectively lower dimension than 2.  The effective reduction we find there, by generating a Poisson point process on the forest patches only, is from $D=2$ to approximately $D=1.66$, see Subsection \ref{Sec:PoissonForest} for more details. We could therefore conclude, that even when finding $\beta\approx0$ for the fit, this may not yet indicate a complete absence of repulsion.

%%%%%%%%%%%%%%%%%%%%%%%%%%%%%%%%%%%%%%%%%%%%%%%%%%%%%%%%%%%%%%%%%%%
\subsection{Interaction among two species}\label{Sub:interact}

\begin{center}
	\begin{figure}[h]
		\begin{minipage}{0.02\linewidth}
			\begin{turn}{90}\hspace{5pt}\textbf{Eagle Owls (P)} \hspace{10pt} \textbf{Eagle Owls (F)}\end{turn}
		\end{minipage}
		\begin{minipage}{0.52\linewidth}
		\hspace{0.0\linewidth} \textbf{vs. Common Buzzards
		} \hspace{0.1\linewidth} \textbf{
		vs. Goshawks} \hspace{0.0\linewidth} \hspace{0.1\linewidth}\\
		\includegraphics[width=0.48\linewidth,angle=0]{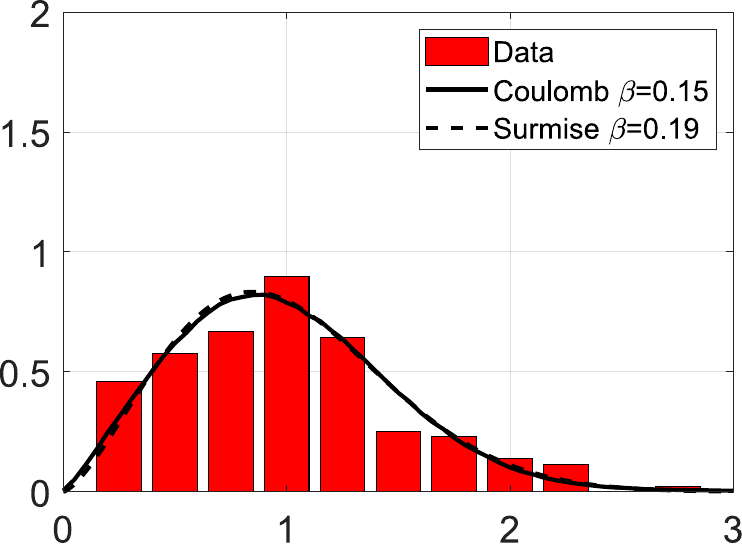}
		\includegraphics[width=0.48\linewidth,angle=0]{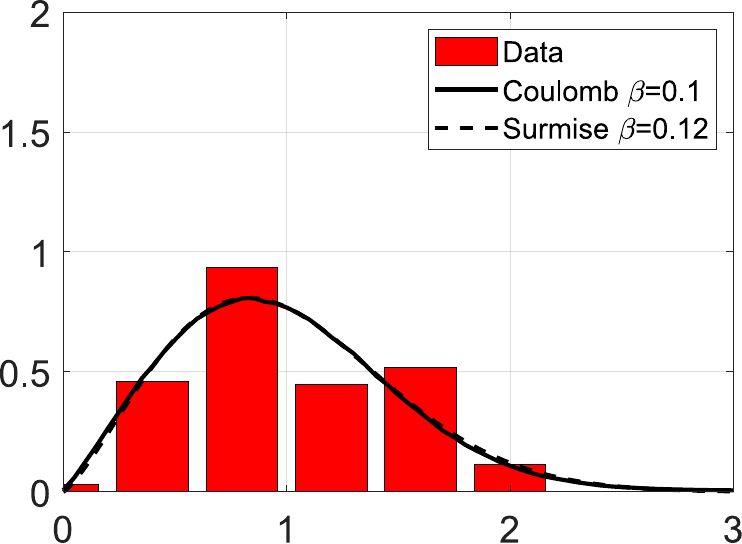}\\

		\includegraphics[width=0.48\linewidth,angle=0]{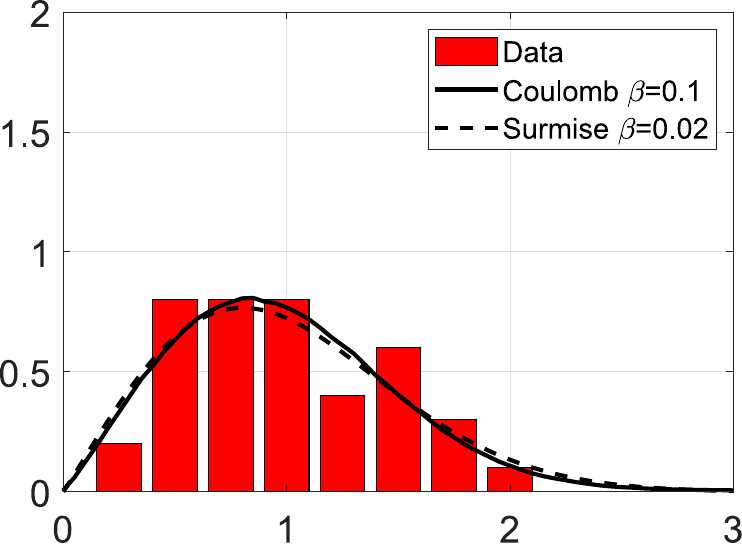}
		\includegraphics[width=0.48\linewidth,angle=0]{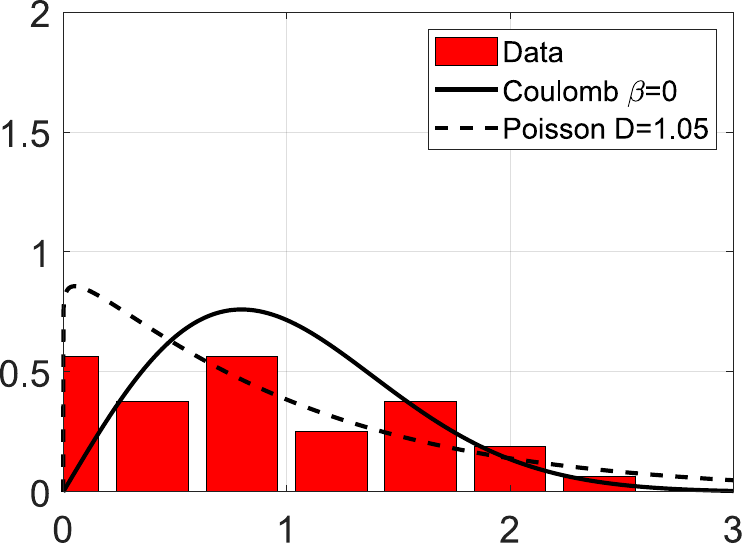}\\
		\centering
		\end{minipage}
		\begin{minipage}{0.02\linewidth}
		\begin{turn}{90}\hspace{5pt}\textbf{Goshawks} 
\end{turn}	\end{minipage}
		\begin{minipage}{0.31\linewidth}
			 \textbf{			 vs. Common Buzzards} \vspace{2pt}\\
			\includegraphics[width=0.96\linewidth,angle=0]{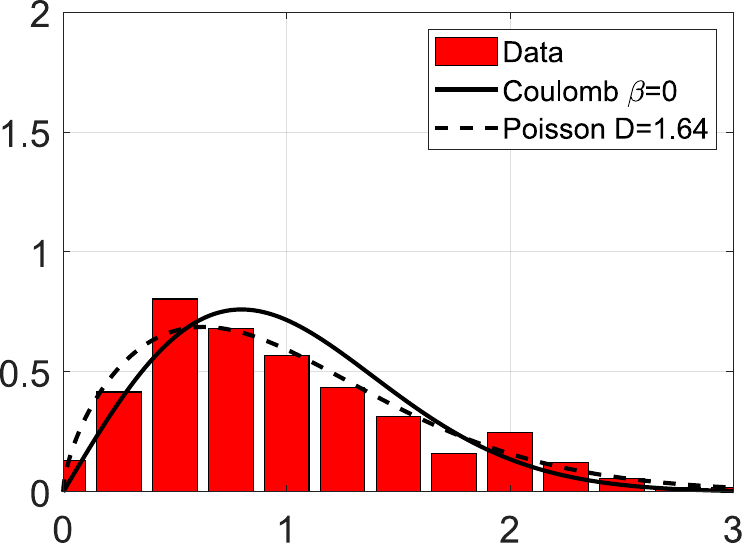}\\
			\centering
		\end{minipage}
		\caption{2D Coulomb gas and surmise fits for the $\beta$-values of the NN spacing distributions (full respectively dashed line) between all pairs of  species for all years, 
	%%%%%%%
{	
		cf. Table \ref{tableNumbers}, where for Eagle Owls we distinguish (F) and (P) in the top row and bottom row, respectively. 
For the comparison with Eagle Owls (F) we have 174 spacings (top left and middle), 
and for Eagle Owls (P) we have 40 spacings (bottom left and middle). In the comparison Goshawk to Buzzard (right column) we have 423 spacings available.  
}		
%%%%%%%%%%%%%%		
When the best fit to the 2D Coulomb gas gives $\beta=0$ (bottom middle and right), we  fit instead to the Poisson distribution with $D<2$ from Eq. \eqref{Eq:poiD} (dashed line), with the resulting value for $D$ given in the inset.}
		\label{Fig:Interspecies}
	\end{figure}
\end{center}

In this subsection we quantify the repulsion between all combinations of two different species of birds, while keeping the two groups of Eagle Owls (F) and (P) separate, see Figure \ref{Fig:PlotBirds} left. Following the 2D Coulomb gas analogy, it would be perhaps natural to associate to each species a different charge, according to their observed (average) repulsion strength. However, such a multi-component 2D Coulomb gas would depend on the ratio of the different charges, which may change over the years.  
If one charge is very abundant, we may consider the others to be screened, but this is also not always the case. The multi-parameter fit is therefore not easy to do, and we stick to the simple one-parameter fits for each pair of species instead, see Figure \ref{Fig:Interspecies}. We thus fit one $\beta$-value to the spacings between species $A$ and $B$.

In order to avoid an over counting of spacings, we always go through the number of points (nest) of the less abundant species between the two, and find its NN to the more abundant species in a given year, defining an ensemble. 
Because in all years the number of pairs of Common Buzzards is larger than the pairs of Goshawks, which in turn is larger than the number of pairs of Eagle Owls ((F) or (P)), the ordering is clear. For example, for the Common Buzzard - Goshawk interaction we go through all Goshawk nests in a given year and find its NN Common Buzzard nest. 
Our statistic is thus limited by the population of Goshawks or Eagle Owls, respectively. 
For that reason, we choose to make an average over all ensembles of 21 years in this subsection. 
%%%%%%%
{
The respective sums of all spacings per species over the entire period can be found in Table \ref{tableNumbers}, last row.}		
%%%%%%%%%%%%%%		
Furthermore, we restrict ourselves to NN spacings only. 
The unfolding of the data has been made using all occupied nests from all three species per year, to get an approximate mean global density. This certainly represents a simplification, but does not introduce any extra bias.

The following picture emerges from Fig. \ref{Fig:Interspecies}. All fitted values are closer to $\beta=0$, with the largest repulsion observed between Common Buzzards and Eagle Owl (F) at $\beta=0.15$. Thus the 2D Coulomb gas and surmise values are very close. 
The NN spacing between Goshawks and Eagle Owls (P) is not well fitted by Poisson at $D\leq 2$, 
nor by a 2D Coulomb gas at $\beta>0$.

Because we average over all years, we have to compare with the values in Fig. \ref{Fig:GroupingYears} right column, for the repulsion within each species for all years: $\beta=0.6$ for Common Buzzards, $\beta=0.8$ for  Goshawks, and $\beta=0.5$ for Eagle Owls (F). 
Clearly, the repulsion measured in $\beta$ is much weaker between different species than within one species. 
The strongest interspecies repulsion measured in this way is between Common Buzzards and Eagle Owls (F) with $\beta=0.15$, followed by that between Goshawks and Eagle Owls (F), and Common Buzzards and Eagle Owls (P) at $\beta=0.1$ each. Do these low values of $\beta$ close to zero indicate, that there is almost no repulsion between different species, as in a random Poisson point process in 2D? As already mentioned in the previous subsection, we investigate in Subsection \ref{Sec:PoissonForest} if the random point process on the forest patches reduces the dimension, finding $D=1.66$ as effective dimension. 
This seems to indicate that even low $\beta$-values close to zero  represent a certain repulsion. For the interaction between Common Buzzards and Goshawks we (perhaps coincidentally) find that the best fit is obtained by the Poisson distribution \eqref{Eq:poiD} close to this effective dimension at $D=1.64$, and in that sense there is no repulsion also according to this measure between the two. 
The plot with the lowest statistics between Goshawks and Eagle Owls (P) could be interpreted in the same way, beside the low quality of the fit. 

The finding of a small repulsion between the three species is in line with empirical findings that intraspecific competition is commonly stronger than interspecific competition in avian predators \cite{Krueger2002a,Krueger2002b}. The effect of Eagle Owls is rather different and has been shown to be clearly negative for both Common Buzzard and Goshawk \cite{CBK,MCHK}. These interactions are, however, very dynamic at a very small scale \cite{CBK,MCHK}, and hence the 
%simple models 
%%%%%%%
{
simplistic 2D Coulomb gas picture 
employed here is perhaps at its limits in detecting these interactions.}

%%%%%%%%%%%%%%%%%%%%%%%%%%%%%%
\section{Methodology - Variations of the random point process} \label{Sec:Method}
%%%%%%%%%%%%%%%%%%%%%%%%%%%%%%%%%%%%%%%%%%%%%%
\subsection{Poisson point process in varying dimension $D$}\label{Sec:PoissonForest}

In this section we investigate if the 2D Poisson point process used so far indeed describes the situation of nests placed as independent random variables in the plane. 
It has been observed that all three species of birds of prey invariably breed in forest patches. 
A look on Fig. \ref{Fig:PlotBirds} right thus poses the question, if the forest represents a lower dimensional,  fractal domain in the full two-dimensional area. Notice that we completely ignore the elevation of the terrain, by treating it as two-dimensional. It varies  from about 70 m to 300 m in height above sea level, compared to the dimension of roughly $12\times25$ km extent of the monitored area.

We will try to answer this question by generating a Poisson point process solely on the forested area, determine its NN spacing distribution numerically, and compare it to the analytic result for the NN spacing distribution of a Poisson point process for general dimension $D$, by fitting $D$ as a free parameter. This distribution is well known for integer dimension $D$, see e.g. \cite{SaRibeiroProsen}, and we simply analytically continue to real $D>0$ here:
\begin{equation}
\label{Eq:poiD}
	p_{\rm Pois, D}^{\rm (NN)}(s) = D \left(\frac{\Gamma(1/D)}{D}\right)^D s^{D-1} \exp\left[-\left(\frac{\Gamma(1/D)}{D}\right)^D s^D\right]\sim s^{D-1}\ .
\end{equation}
It is normalised with first moment equal to unity. The repulsion $\sim s^{D-1}$ originates entirely from the (fractional) area measure.

\begin{center}
	\begin{figure}[b]
		\centering
		\includegraphics[width=0.45\linewidth,angle=0]{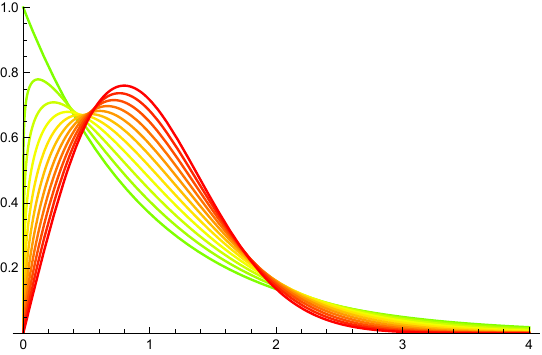}
		\includegraphics[width=0.4\linewidth,angle=0]{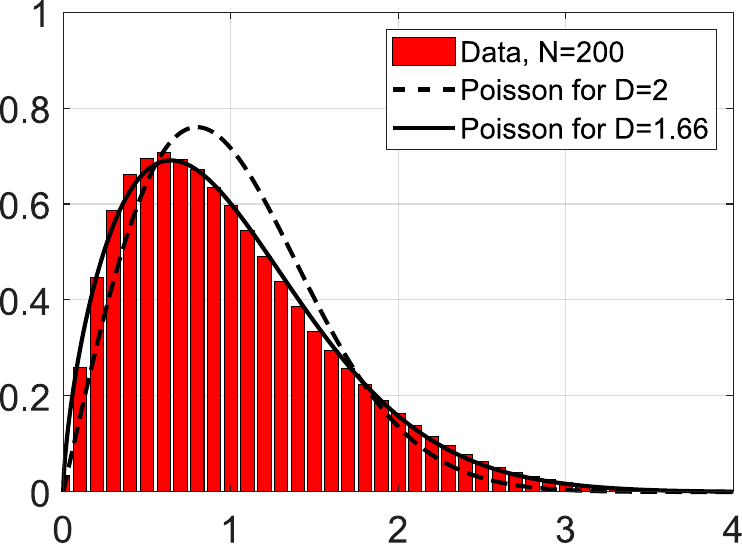}
		\caption{
{\bf Left}: NN spacing distribution of the Poisson point process \eqref{Eq:poiD} for various dimensions from $D=2$ (red) down to $D=1$ (green), in steps of 0.1. The maximum is moving from right to left when going from $D=2$ to $D=1$.
{\bf Right}: Fit of the dimension $D$ in \eqref{Eq:poiD}	(solid curve) to the random 
point process generated on the forested area as described in the main text (histograms). 	From the fit we obtain an effective dimension of $D=1.66$. For comparison, the NN spacing distribution in $D=2$ is also shown (dashed curve).}
		\label{Fig:PoissonForest}
	\end{figure}
\end{center}

The Poisson point process on the forest is generated as follows. A certain number $N$ of points is distributed independently in the monitored area. If they fall onto a green forest patch in Fig. \ref{Fig:PlotBirds} right, they are accepted, else they are rejected. Note that we accept all  areas with trees, i.e., not just the main forest depicted as a green band in Figure \ref{Fig:PlotBirds}. 
In the limit $N\gg1$ with a fixed area, we would recover a collection of two-dimensional patches, given the number of points per patch is large enough.  In order to 
capture roughly the same length scale as the distances between occupied nests, we stop after having $N=200$ accepted points, which is about the number of Common Buzzard pairs in the entire area.  
For the fit we use a Kolmogorov-Smirnov fit to the CDF of eq. \eqref{Eq:poiD}, see Figure \ref{Fig:PoissonForest} for the result.

We find that the Poisson point process generated as described indeed leads to slight reduction of dimension from $D=2$ to an effective dimension of $D=1.66$. Consequently, we may conclude that a fit of data to a Coulomb gas with $\beta=0$, or equivalently the Poisson point process with $D=2$, still reflects a small repulsion, compared to the process on the forested area.
Consequently this makes the repulsion found for small $\beta$ more pronounced.

In Subsection \ref{Sub:interact}, where the repulsion between different species is quantified by fitting to the NN spacing of the 2D Coulomb gas with $\beta\geq0$, we encounter the situation that the fit to $\beta=0$ is apparently still not satisfactory, because the maximum of the NN data is further to the left. In that cases we rather fit the dimension $D$ of the Poisson NN distribution \eqref{Eq:poiD}, that has a maximum further to the left, see Fig. \ref{Fig:Interspecies} and the discussion there.

%%%%%%%%%%%%%%%%%%%%%%%%%%%%%%%%%%%%%%%%%%%%%%%%%%%%%%%%%%%%%
\subsection{Varying correlation length: 2D Yukawa interaction}\label{Sec:Yukawa}
In the previous Section \ref{Sec:Interact} we have seen that fitting $\beta$ independently to the NN and NNN spacing distribution may lead to different values. This indicates that the repulsion between the nests cannot always be described by a 2D Coulomb interaction at a single inverse temperature $\beta$. 
For instance, when finding $\beta_{\rm NN}>\beta_{\rm NNN}$ we expect an interaction weaker than Coulomb, or of shorter range. Likewise, for $\beta_{\rm NN}<\beta_{\rm NNN}$ we expect an interaction stronger than Coulomb, or of longer range on that scale.

This motivates us to study a Coulomb-like interaction in this subsection, where the interaction range can be varied: The Yukawa potential $V_{\rm Yukawa}$ in $D=2$ dimensions. We note in passing that the Yukawa interaction arises in particle physics from the scattering of {\it distinguishable} Fermions (points) in the non-relativistic limit, see \cite{PS} for a standard work on quantum field theory, which we follow here for the derivation.

The Yukawa potential can be defined in $D$ (integer) dimensions\footnote{In most textbooks only $D=3$ is considered, including \cite{PS}.} by taking the inverse Fourier transformation of the propagator with mass $m$ and coupling constant $g$:
\begin{equation}
	V_{\rm Yukawa}^{(D)}(\vec{x}) = \int \frac{dq^D}{(2\pi)^D} \frac{-g^2 }{|\vec{q}|^2 + m^2}e^{i\vec{q}\cdot\vec{x}} ,\quad \vec{x}\in\mathbb{R}^D.
	\label{VYukD}
\end{equation}
The integral can be performed using polar coordinates, and in $D=2$ we obtain
\begin{eqnarray}
V_{\rm Yukawa}^{(2)}(\vec{x})
	&=& -\frac{g^2}{4\pi^2} \int_{0}^{\infty} dq q \int_0^{2\pi} d\theta \frac{e^{iqr\cos\theta}}{q^2 + m^2}\nn\\
	&=& -\frac{g^2}{2\pi} \int_{0}^{\infty} dq q\frac{J_0\left(qr\right)}{q^2 + m^2}\nn\\
	&=& -\frac{g^2}{2\pi} K_0\left(mr\right)\ .
	\label{VYuk2}
\end{eqnarray}
It only depends on the radial distance $r=|\vec{x}|$. Here, we used the standard integral representation of the Bessel function of the first kind $J_0(y)$, and $K_0(y)$ 
denotes the modified Bessel function of the second kind. The last equality follows from \cite[6.532]{Gradshteyn}. It has a logarithmic singularity at the origin, 
\begin{equation}
K_0\left(mr\right)\sim -\log(mr/2) - C, \quad \mbox{for} \ mr\to0, 
\end{equation}
where $C\approx 0.577$  is the Euler--Mascheroni constant, 
and vanishes exponentially for large distance as
\begin{equation}
K_0\left(mr\right)\sim \sqrt{\frac{\pi}{2mr}}\ e^{-mr}, \quad \mbox{for} \ mr\to\infty. 
\end{equation}
If we define a length scale by $\gamma=1/m$, we obtain a one-parameter deformation of the 2D Coulomb interaction.  Namely, for fixed distance $r$ and large scale $\gamma\gg1$ ($m\ll1$) we are back to the logarithmic repulsion (shifted by a constant). At fixed $\gamma$, however, the interaction range of the 2D Yukawa potential is much shorter and decays exponentially, while still being logarithmic at short distances. 

\begin{figure}[b]
	\centering
	\hspace{0.1\linewidth} \textbf{Potential} \hspace{0.15\linewidth} \textbf{Nearest Neighbour} \hspace{0.06\linewidth} \textbf{Next-to-Nearest Neighbour} \hspace{0.08\linewidth}\\
	\includegraphics[height=157pt]{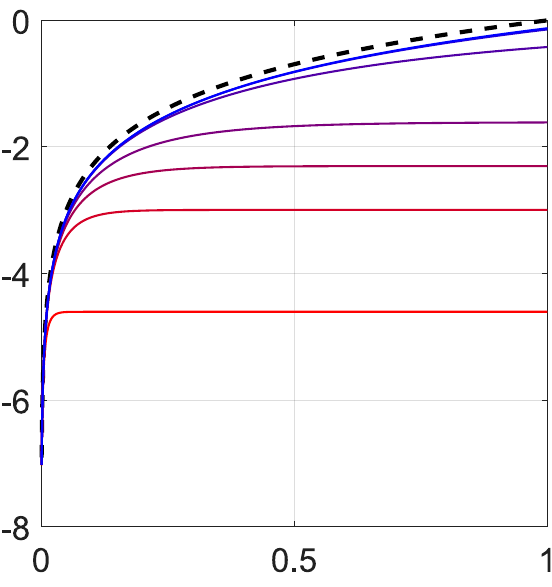}
	\includegraphics[height=157pt]{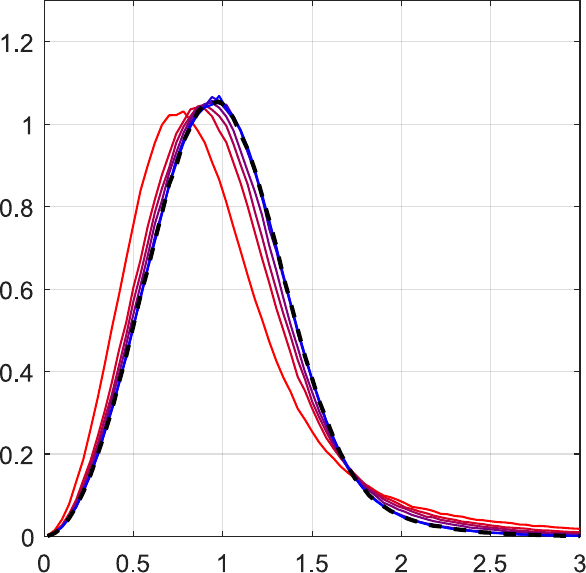}
	\includegraphics[height=157pt]{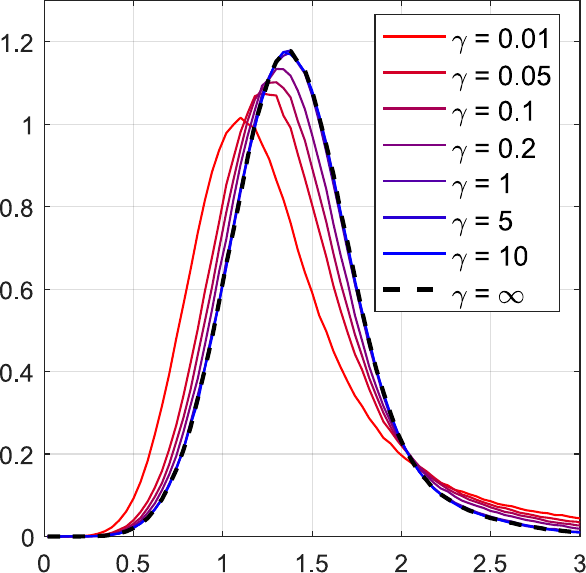}\\
	Distance
	\caption{{\bf Left:} Comparison of the 2D Coulomb potential ($\gamma=\infty$, dashed)  and the the 2D Yukawa potential \eqref{VYukfinal} varying from $\gamma=0.01$ (bottom red full line) to $\gamma=10$ (top blue full line), see also inset in the right plot for the colour coding.
	\textbf{Middle and right:} numerically determined NN and NN spacing distributions, respectively, for the same values of $\gamma$. The maximum increases with $\gamma$ and moves from left to right towards the rightmost curve, given by the 2D Coulomb potential.
	}
	\label{Fig:Yukawa}
\end{figure}

In order to match with the point process \eqref{Eq:Coulomb} of the 2D Coulomb gas at a given inverse temperature $\beta$, we use the following shifted potential \eqref{VYuk2}:
\begin{eqnarray}
	V_{\rm Yukawa}^{(2)}(r) &=& -\beta \left(K_0\left(r\gamma^{-1}\right) - \log(2\gamma) +C\right) \ ,
	\label{VYukfinal}
\end{eqnarray}
identifying  $\frac{g^2}{2\pi}=\beta$. A comparison between the two potentials at fixed $\beta=1$ is shown in Figure \ref{Fig:Yukawa} left.

The NN (NNN) spacing distributions shown in Fig. \ref{Fig:Yukawa} middle (right) have to be determined numerically again, as it was done for the 2D Coulomb interaction in \cite{Buzzard}. That is, we use a Metropolis-Hastings algorithm \cite{Metropolis, MetropolisHastings} with \eqref{VYukfinal} as the potential. The points are initialised independently, and then perturbed iteratively. Each perturbation is always accepted if it leads to lower energy, and accepted with probability $e^{-(V_{\rm after} - V_{\rm before})}$ if it leads to higher energy, otherwise it is rejected. (Note that by including $\beta$ in Equation \eqref{VYukfinal}, we have made it dimensionless.) After a number of iterations, the points are considered to be a sample of the potential. On average, each point is perturbed 100 times. This is considered enough, as increasing the amount of iterations does not change the NN spacing distribution. Unfolding is necessary when computing the NN spacing distribution of the Yukawa potential for $\gamma<\infty$, as the global spectrum is non-uniform.

While the potential $V_{\rm Yukawa}^{(2)}(r)$ differs considerably for the parameter values chosen, the spacing distribution converges rather rapidly to the one of the Coulomb potential at $\gamma\to\infty$, in particular for NN. The reason is that it probes local correlations among the points (which are closer for NN than for NNN), which seems to be rather robust under deformations of the potential. As an aside, such a universality has been proven rigorously in 1D  for all fixed values of $\beta$ for deformed potentials that behave only locally as a logarithm \cite{Martin}.

The fitted value of the scale $\gamma$ could in principle be translated into an actual correlation length, but the effect of lowering the parameter $\gamma$ is qualitatively similar to lowering $\beta$, and only becomes noticeable at relatively small values of $\gamma$ for the NN. Additionally, in order to compare our data with the spacing distribution we first need to unfold. This makes a translation back into a real correlation distance that may depend on the terrain (local density) rather cumbersome. The two-parameter fit with the Yukawa potential does not reconcile the discrepancy between different $\beta$-values obtained for the NN and NNN distributions, and we therefore did not pursue the approach further.
In other words, adding a length scale does not allow us to fit both NN and NNN with the same parameter values. Recall that the actual value of $\beta$ does not have a biological meaning. It rather serves as a relative measure in comparing different species, and the sole purpose of adding a length scale was to compensate for differences in the fitted $\beta$ for NN and NNN distributions. Fitting both NN and NNN at the same time would not improve this, as we would merely observe an average of the two distributions. Because of the relatively small amount of available data, especially for the Eagle Owls, it is probably not correct to weight the contributions from NN and NNN equally in determining $\beta$ and $\gamma$. However, any other choice of weighting them easily becomes arbitrary.

%%%%%%%%%%%%%%%%%%%%%%%%%%%%%%%%%%%%%%%%%%%%%%5
\subsection{Independence of points: Reuse of nests}\label{Sec:Reuse}

One of the key assumptions in deriving the Poisson NN and NNN spacing distributions in \eqref{PoiNN} and \eqref{PoiNNN} is the independence of points in this process. In this subsection we investigate if in the absence of correlation (repulsion), which we found in some data sets by agreeing with Poisson, it is justified. The reason for this question is that some of the nests are reused, not only within one species but also between different species, and we will present data about this fact below. Such an analysis is possible because all old and new nests were marked precisely with GPS coordinates, up to an error of about 10 m.
%%%%%%%%%
{ In this subsection we do not split the group of Eagle Owls into two.}

\begin{figure}[h]
	\centering
	\hspace{0.02\linewidth} \textbf{Buzzards} \hspace{0.18\linewidth}
	\textbf{Goshawks} \hspace{0.17\linewidth}
	\textbf{Eagle Owls}\\
	\begin{turn}{90}\hspace{30pt}\tiny Probability of Reuse\end{turn}
	\includegraphics[width=0.3\linewidth,angle=0]{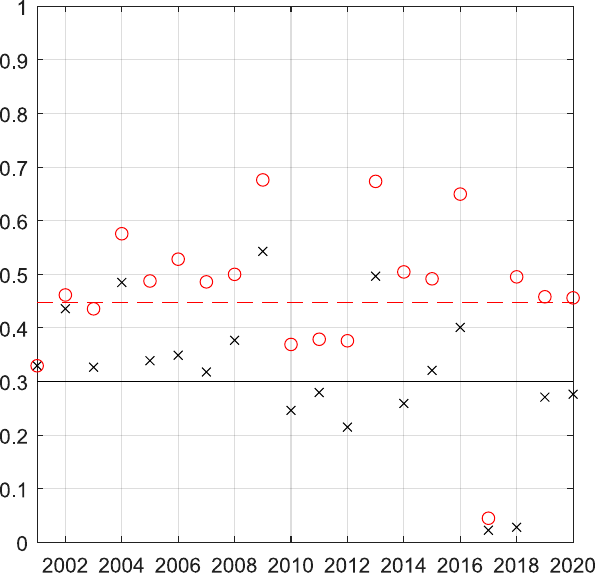}
	\includegraphics[width=0.3\linewidth,angle=0]{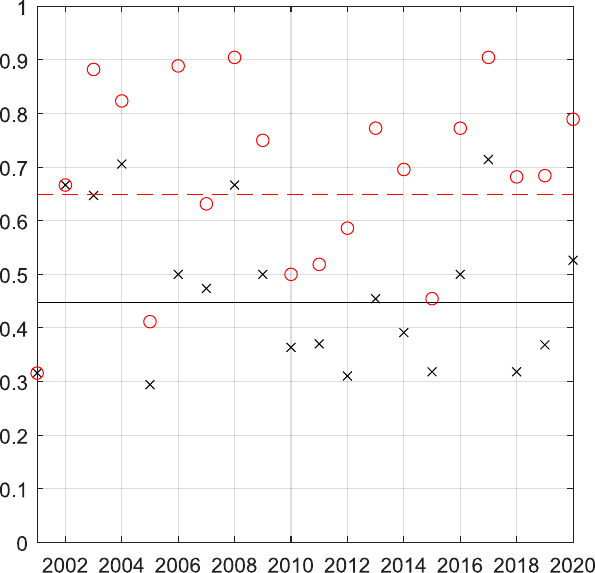}
	\includegraphics[width=0.3\linewidth,angle=0]{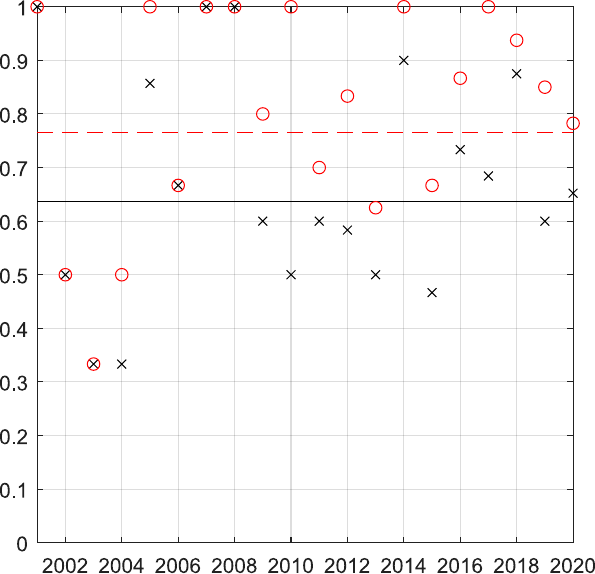}
	\caption{Percentage  of nests that have been reused per year by each species. The black crosses indicate the fraction of nests that were reused  from the previous year by the same species, i.e. Common Buzzards reusing a Common Buzzard nest in the left plot. The black full line indicates the mean over the entire period of observation. 	
The red circles indicate the fraction of nests that one species reused from all nests occupied by any of the 3 species in any previous year. i.e. a Common Buzzard reusing a nest from a Common Buzzard, Goshawk or Eagle Owl from any previous year in the left plot. 
The red dashed line again indicates the mean value. Notice that both percentages may sometimes coincide,
%%%
{ as observed for the Eagle Owls.}
	}
	\label{Fig:Reuse}
\end{figure}

The amount of reusage of occupied nests from one year to the next year varies between 30\% (Common Buzzards), 45\% (Goshawks) and 65\% (Eagle Owls (F)+ (P)) within one species, on average. The reuse of any nest from any previous year is about 10-20\% higher. 
The frequent reuse of nest sites over years and even decades has been documented previously for the study area and is typical of avian predators \cite{Krueger2002b}. Habitat heterogeneity in many dimensions is commonly observed in these species \cite{KL01}, and individual preferences for certain habitat features are also common \cite{Krueger2002b}, leading to frequent reuse of nests. In other species, the same nest has been in use for centuries \cite{Ellis09} and even millennia \cite{BBN09}.

What is the consequence for our data analysis? In Section \ref{Sec:Interact} we treated every set of occupied nests per year as an independent ensemble, in order to compare it to a statistical mechanics 
%model, 
%%%%%%%
{ picture,} 
that represents an equilibrium configuration.  
The above data show that this is a simplification, that is only partly justified. The fact that a substantial fraction of nests is reused, both for the repulsion within one species or among two different species, means that not all NN and NNN positions are new. 

In order to improve our statistics we have then made an ensemble average over 10 consecutive years (in 
\cite{Buzzard} over 5 years). We have not analysed if after two or more years the positions become more independent or randomised, but it seems reasonable to assume that. Our situation could be compared to ensemble generation in statistical mechanics. If a new configuration is generated from an old one, e.g. with the Metropolis--Hastings algorithm used to generate the 2D Coulomb and 2D Yukawa ensembles in this paper, one has to monitor the autocorrelation time before a new configuration can be accepted as independent. In the generation of points, we could simply let enough iterations go by, but for the observed data we cannot wait one or more years, until the next set of occupied nests is more independent, without losing much information (and statistics). This is because the repulsion within one or more species can change over time, as we have observed.

There are ecological reasons 
why certain locations of nests are more popular than others, which is why they get reused. Plus it takes time and effort to build a new nest.
Certain nest sites are associated with higher reproductive success, independent of the individuals using these nests \cite{Krueger2002b}. The proximity of high quality hunting habitat, less exposure to predators and parasites, and less disturbance from humans are just some of the key factors \cite{Krueger2002b,CBK}.

The fact that there is a large number of used nests in the observed area seems to guarantee at least to some extent, that a kind of  "random sampling" can take place between consecutive years.
There is an estimate on the total number of 500-600 reusable nests in the area for birds of prey.

As a final remark, there seems to be no correlation between the population growth and the percentage of reusage of nests.
%%%%% %%%%%
{
It is approximately constant in Fig. \ref{Fig:Reuse} and largely fluctuates around the mean. As we have seen in Fig. \ref{Fig:AllBirds} 
(and Table \ref{tableNumbers}) for the Common Buzzard,  their population has grown from 63 to 267 pairs, 
and from a single pair to 23 for all Eagle Owls in total, see Fig. \ref{Fig:EagleOwlGroup} and Table \ref{tableNumbers}.} 
%%%%%%%5
The Goshawk population has also grown but more slowly, and then declined again. 
Together with the high number of empty nests available each year this could be argued to be in favour of a certain independence of the locations of nests each year, despite the high fraction of reusage. 

\newpage
%%%%%%%%%%%%%%%%%%%%%%%%%%%%%%%%%%%%%%%%%%%%%%%%%%
\section{Dicussion and Open Questions}\label{Sec:Conclusio}

%%%%%%5
{ 
This study has shown that a simple, one-parameter dependent approach from statistical mechanics, without any biological assumptions, 
%background, 
%nevertheless has surprisingly well explained 
has the potential to explain important features in this three-species guild of avian predators.} 
%%%%%
We have shown that the intraspecific repulsion measured through $\beta$ clearly deviates from Poisson statistics. The relative ranking in repulsion strength amongst the three species, averaged over all years, has been found compatible with experience from ecology. 

The correlation between $\beta$ and the time dependence of the population have been found plausible for at least two of the three species analysed. For the Common Buzzard the steep rise in population from 
%%%%%
{
63 to 267 pairs} 
%%%%%%
has been paralleled by a clear rise in $\beta$, in particular in the NNN spacing, showing also an  increased interaction range. For the Goshawk the population has been more stable around 20 pairs, with a decline after a peak around 10 years ago. Both NN and NNN show a clear decrease in the fitted $\beta$. In contrast for the Eagle Owls, with the population increasing from a single pair to over 
%15, 
23,
we have seen a steep decrease in $\beta$ for NN and no repulsion between NNN. This relation is not clear to us and may be due to the limitation of 
%the model, 
%%%%%%%
{ our approach,} 
or an artefact due to very low statistics. 
The change in $\beta$ over time might hence be a fruitful further topic for investigation.

At the same time, the results also clearly indicate some limitations when it comes to interspecific interactions. 
We have found the repulsion between Eagle Owls in the forest (F) on the one hand, and the Common Buzzard respectively Goshawk on the other hand to be significantly smaller than the respective intraspecific repulsion, which is plausible. Both interspecies repulsions have values of $\beta=0.15-0.1$ above Poisson at $\beta=0$. Taken together with the effect of the terrain, effectively lowering the dimension of the Poisson distribution, this makes the repulsion found more pronounced and significant. 
For the Eagle Owls in the plains (P) the same order of repulsion was only observed with the Common Buzzard, and no repulsion was seen for the Goshawk, neither among the latter two. 
In view of the continuing growth of population in Common Buzzards and Eagle Owls, notably populating the plains for the latter, and the population of Goshawks being under pressure, this seems to be a limitation of the %model 
%%%%%%%
{ our approach (and perhaps present statistics).}

%From the point of view of the statistical mechanics, 
%model, 
We have shown that a simple formula based on a surmise from random matrix theory allows to quantify and discuss the intra- and interspecies NN repulsion as well, instead of using the numerically generated spacings distributions from a full 2D Coulomb gas, that also include NNN.  Methodologically, we have quantified the effect of the terrain. In particular, it is not responsible for the repulsion observed. We have also discussed the 2D Yukawa interaction and spacing ratios as alternative %models. 
%%%%%%%
{ measures.}
The extensive reusage of nests, especially among Eagle Owls, has put the assumption of independence of ensembles as realised by consecutive years under scrutiny. However, at least from a random matrix perspective we know that there, spectral statistics shows a certain robustness or universality under such perturbations.
%%%%%%
{
Finally, as a cross check we have made a numerical experiment for two examples of year with medium to low statistics. Introducing a random displacement of the original locations of the nests by an amount comparable to the mean spacings between the nests, the displacement destroys the repulsion we have seen in fitting $\beta$ to the NN spacing, and rapidly leads to a Poisson distribution.

It would be very valuable to find an underlying microscopic model that leads at least approximately to the kind of spacing distribution of a 2D Coulomb gas, as the one we have applied in our comparison with the data.}
%%%%%%%%%%%%%%%

We feel that this example of %model 
transfer between disciplines has been an insightful exercise.
% for both disciplines involved. 
It remains to be seen whether interspecific interactions might be better captured with another decade of data with increased statistics, 
and with all three species at their ecological carrying capacity.\\ 

{\it Acknowledgements:} We acknowledge 
funding by the German Research Foundation (DFG)
 through grant CRC 1283/2 2021 "Taming uncertainty and profiting from randomness and low regularity in analysis, stochastics and their applications"  %317210226 
 (GA and PP), and CRC/Transregio 212 "A Novel Synthesis of Individualisation across
Behaviour, Ecology and Evolution: Niche Choice, Niche Conformance, Niche Construction (NC$^3$)" (OK). 
AM would like to thank Kurt Mielke for improving the speed of the Metropolis-Hastings code and providing a server for it, which made both the many parameter combinations of the Yukawa potential and  larger $N$ feasible. We thank Michael Baake for useful discussions and the anonymous referee for suggesting the random displacement test.

%%%%%%%%%%%%%%%%%%%%%%%%%%%%%%%%%%%%%%%%%%%%%%%%%%%%%%%%%%%%%%%%%%%%%%%%%%%
%%%%%%%%%%%%%%%%%%%%%%%%%%%%%%%%%%%%%%%%%%%%%%%%%%%%%%%%%%%%%%%%%%%%%%%%%%%
\begin{appendix}
%%%%%%%%%%%%%%%%%%%%%%%%%%%%%%%%%%%%%%%%%%%%%%%%%%%%%%%%%%%%%%%%%%%%%%%%%%%	

\section{Alternative to Unfolding: Complex Spacing Ratios}\label{Sec:Ratios}

In this appendix we briefly discuss complex spacing ratios as introduced in \cite{SaRibeiroProsen} and compare a subset of our data to several moments of these. However, the outcome is not conclusive, as we will see.

Complex spacing ratios have become a popular tool in analysing two-dimensional data sets, compared to NN or NNN spacing distributions in radial direction. The advantage compared to the latter is that, first, no unfolding is necessary, which is often more difficult in two compared to one dimension. Second, it also gives both angular and radial information about the interaction between the points. It is defined as follows. Suppose we have an eigenvalue at $z_k$, with NN at $z_k^{\rm NN}$, and NNN  at $z_k^{\rm NNN}$, in radial distance. The complex spacing ratio is then defined as 
\begin{equation}
u_k=\frac{z_k^{\rm NN}-z_k}{z_k^{\rm NNN}-z_k}\ .
\label{spacingratio}
\end{equation}
The goal is then to determine the probability distribution $\rho(u=re^{i\theta})=\rho(r,\theta)$ of this spacing ratio, in the limit of a large number of particles $N\to \infty$ in the bulk of the spectrum, for a given point process. For the Poisson process, it is known to be flat on the unity disc \cite{SaRibeiroProsen},
\begin{equation}
\rho_{\rm Poi}(u)=\frac{1}{\pi}\Theta(1-|u|)\ .
\label{Poi-spacing}
\end{equation}
For the complex Ginibre ensemble explicit $(N-1)$-fold integral representation are known \cite{SaRibeiroProsen}, as well as approximate expressions that converge very rapidly \cite{DusaWettig}.  
 
There are several disadvantages in our situation, however. First, the amount of data we have available is rather small - even if very large for biological standards. This makes even a qualitative comparison to 2D plots very hard, if not impossible. 

The difficulty to make 2D-fits was anticipated in \cite{SaRibeiroProsen}, and thus alternatively integrals over the angle or the radius where proposed, 
\begin{equation}
\rho(r)=\int_0^{2\pi}d\theta r \rho(r,\theta)\ ,\quad \rho(\theta)=\int_0^1drr \rho(r,\theta)\ ,
\label{int-spacing}
\end{equation}
and then to consider moments thereof.
Here, we assume that the limiting support of complex eigenvalues is normalised to the unit disc, as in Section \ref{Sec:2DC} for the general 2D Coulomb gas. 
Analytic and approximate formulas have been worked out both for the Poisson case, and the complex Ginibre ensemble, which we reproduce here for completeness. Inserting the 
expression \eqref{Poi-spacing} for the Poisson ensemble into the definition \eqref{int-spacing}, we obtain \cite{SaRibeiroProsen} 
\begin{equation}
\rho_{\rm Poi}(r)=2r\ , \quad \rho_{\rm Poi}(\theta)=1/(2\pi)\ .
\end{equation}
This leads to the following expression for the first moments in the Poisson case:
\begin{align}
\label{momPoi}
	\begin{split}
		\langle r \rangle_{\rm Poi}=& \int_0^1dr\,r \rho_{\rm Poi}(r)	
\ =\ \frac{2}{3}\ ,\\
		\langle r^2 \rangle_{\rm Poi}=&\int_0^1dr\,r^2 \rho_{\rm Poi}(r)	
		\ =\ \frac{1}{2}\ ,\\
		\langle \cos(\theta) \rangle_{\rm Poi} =&
\int_0^{2\pi}d\theta \cos(\theta)\rho_{\rm Poi}(\theta)	
		\ =\ 0\ ,\\	
						\langle \cos^2(\theta) \rangle_{\rm Poi} =&\int_0^{2\pi}d\theta \cos(\theta)^2\rho_{\rm Poi}(\theta)	
				\ =\ \frac{1}{2}\ ,\\
						\langle r \cos(\theta) \rangle_{\rm Poi} =&\ \int_{0}^{2\pi}d\theta\int_{0}^{1}dr r\, r \cos(\theta) \rho_{\rm Poi}(r,\theta) \ =\ 0\ .
	\end{split}
\end{align}
Each moment is normalised. The angular and radial integrals decouple in all cases.
The moments for complex Ginibre are found in \cite{DusaWettig} and reproduced here in Table \ref{tableDW} below for completeness.

\begin{table}[h]
\begin{tabular}{c|r|c}
limiting moment& Ginibre& Poisson\\
\hline
$\langle r \rangle$& 0.739 &2/3\\%66010
$\langle r^2 \rangle$ & 0.581 &1/2\\%84906
$\langle \cos(\theta) \rangle$&  -0.247 &0\\%83082
$\langle \cos(\theta)^2 \rangle$ & 0.450 & 1/2\\%91388
$\langle r \cos(\theta) \rangle$ &-0.189 &0\\%82912
\end{tabular} 
\vspace{5pt}
\caption{Numerically determined moments in the complex Ginibre ensemble from \cite{DusaWettig}, rounded to 3 digits (middle column), and the corresponding values for Poisson from \eqref{momPoi} \cite{SaRibeiroProsen}.}
\label{tableDW}
\end{table}

For the 2D Coulomb gas with intermediate values $0<\beta< 2$, to our knowledge  no such relations have been worked out, in particular if the corresponding values are in between the Poisson and Ginibre values. Furthermore, the effect of the interaction range not being of Coulomb type on the moments is not clear. These are crucial for our quantitative comparison.

\begin{figure}[h]
	\centering
	\hspace{0.01\linewidth} \textbf{Common Buzzards} \hspace{0.30\linewidth}
	\textbf{Goshawks}\\[1ex]
	\includegraphics[width=0.45\linewidth,angle=0]{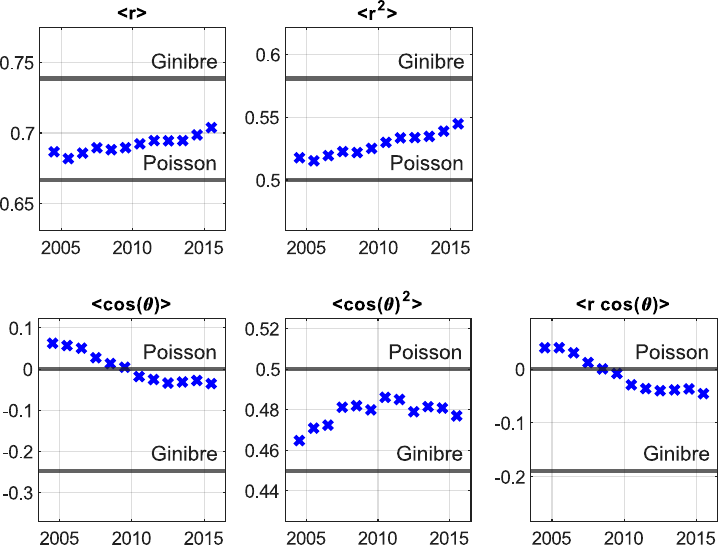}\hspace{30pt}
	\includegraphics[width=0.45\linewidth,angle=0]{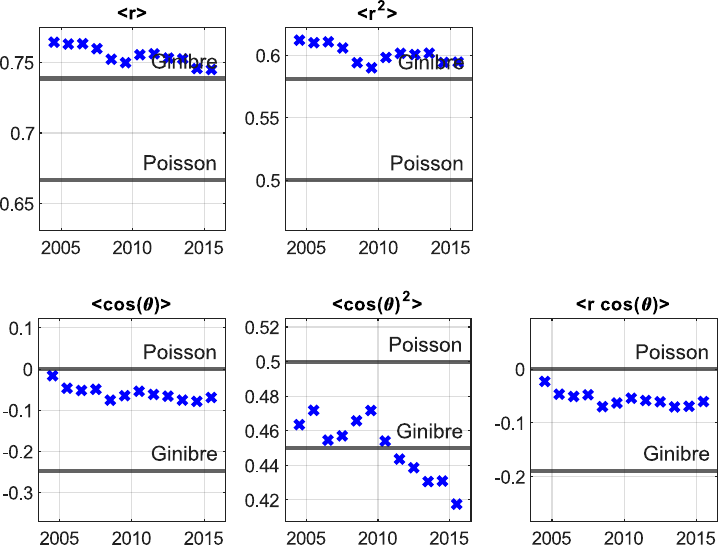}
	\caption{
The moments from complex spacing ratios in \eqref{momPoi} for Poisson and from Table \ref{tableDW} for complex Ginibre \cite{DusaWettig} are compared to the corresponding moments for Common Buzzards (left) and Goshawks (right). For a better comparison we have put the moments where the Ginibre (Poisson) value is higher in the top (bottom) row.}
	\label{Fig:ComplexRatio}
\end{figure}

In Figure \ref{Fig:ComplexRatio} we compare the moments of the complex spacing rations from Poisson and complex Ginibre ensemble to those of the spacings among nests of Common Buzzards (left) and Goshawks (right), for which we have the most data. As in Section \ref{Sec:Interact} we use a time moving average.

In the comparison Fig. \ref{Fig:ComplexRatio} for the Common Buzzards (left) the top row with the radial moments shows a clear tendency over time, where the value for the moments increases from above Poisson towards Ginibre. 
This is consistent with the findings from Section \ref{Sec:indiv}  for the fitted $\beta$-values for NN and NNN spacing distribution for Common Buzzards, see  Figure \ref{Fig:AllBirds} left, where an increase was seen for NN (NNN) from about 0.4 to 0.7 (0.0 to 0.8).  
In contrast, the angular moments in the bottom row are less conclusive. While the left and right plots are close to Poisson and show a decrease towards the negative value for Ginibre, the middle plot is scattered somewhat between Poisson and Ginibre and rather decreasing towards Poisson. 

For the Goshawks the picture is even less clear from the complex spacing ratios.  While the radial moments in the top row show values above Ginibre (which is above Poisson), with a slight tendency going down towards the Ginibre value,  the bottom plots with angular moment (left and right plot) are closer to Poisson, with no clear  trend. The bottom middle plot is quite scattered, starting in between Poisson and Ginibre and then moving clearly below the Ginibre value. 
In comparison,  in Section \ref{Sec:indiv}  for the fitted $\beta$-values for NN and NNN spacing distribution for Goshawks, see  Figure \ref{Fig:AllBirds} right, a clear decent from $\beta$-values from about 1.5 down for 0.5 for both NN and NNN was detectable. 

In the comparison of $\beta$-values within one species and amongst different species, it was important to have even approximate values for $\beta$, that quantify the repulsion. With the predictions from moments of complex spacing rations such a quantitative comparison does not seem to be easily possible, at least not with the predictions we have at hand. Second, we cannot exclude an influence of the spherical geometry on the predicted complex spacing ratios, a situation we do not have for our observed data. This seems to be indicated by the unclear trend form the angular momenta. The spacing distributions we use in the main text are local quantities which a priori seem to be less sensitive to the geometry.

\newpage
%%%%%%%%%%%%%%%%%%%%%%%%%

\section{Comparison surmise and 2D Coulomb gas NN spacing for small $\beta$}\label{Sec:surmise}

\begin{figure}[b]
	\centering
	\includegraphics[width=0.99\linewidth,angle=0]{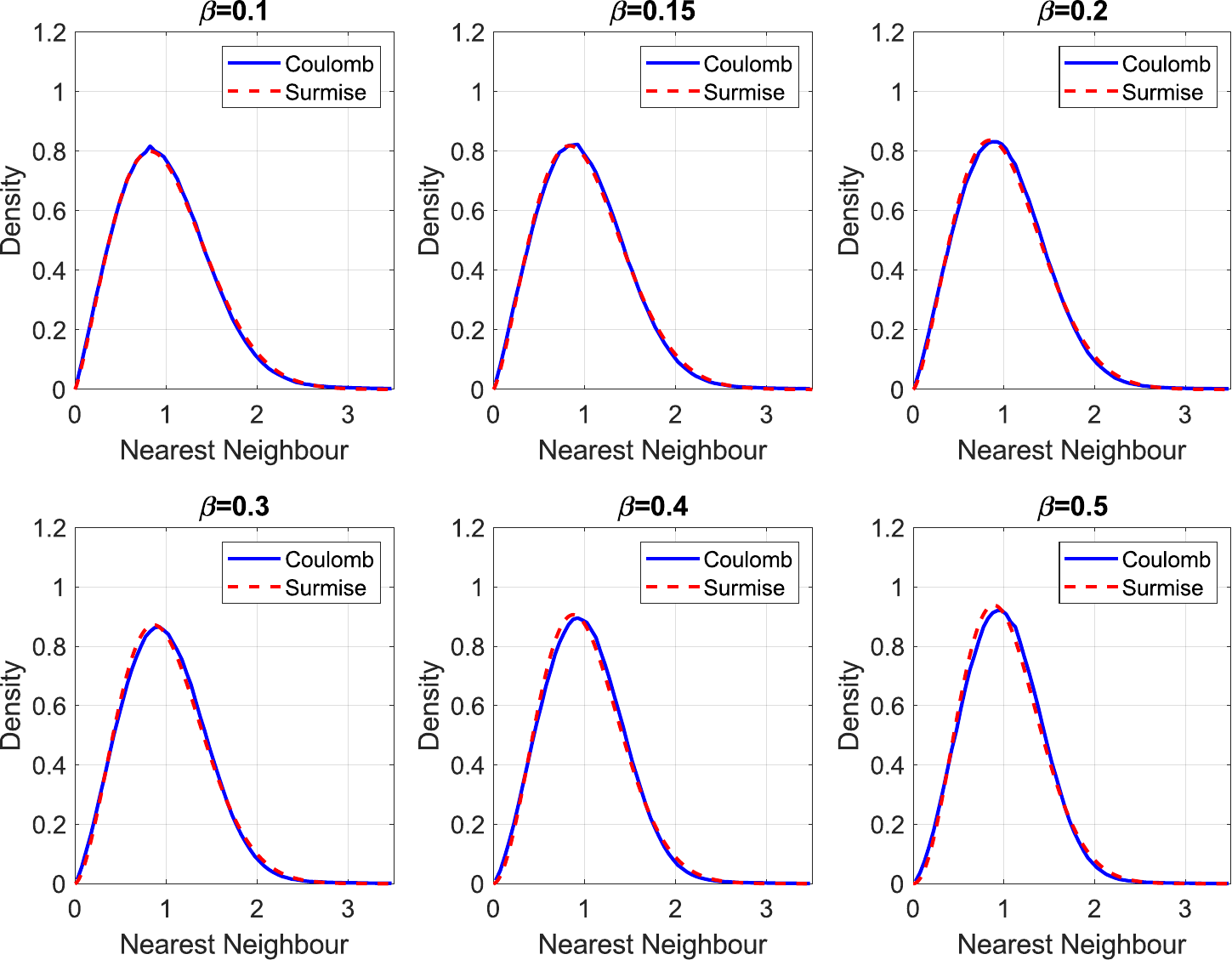}\hspace{30pt}
	\caption{
Comparison of the surmise \eqref{Eq:betasurmise}  with \eqref{Eq:betafit} (dashed red line), and the numerically determined NN spacing distribution from the 2D Coulomb gas \cite{AKMP} (full blue line).}
	\label{Fig:surmise-Coulomb}
\end{figure}

In this appendix we illustrate the deviation between the numerically determined NN spacing distribution from \cite{AKMP} and the surmise \cite{AMP} given in \eqref{Eq:betasurmise} together with \eqref{Eq:betafit}, for small values of $\beta$. The deviation quantified in terms of the standard deviation and the Kolmogorov-Smirnov distance can be found in \cite[Table 1.]{AMP}. 

The fits in $\beta$ from the 2D Coulomb gas to the data presented in Section \ref{Sec:Interact} are done in steps of $0.1$ in $\beta$, except for an intermediate step at $\beta=0.15$ for small $\beta$, see Fig. \ref{Fig:surmise-Coulomb} top middle. In contrast, the fits to the surmise in Section \ref{Sec:Interact} are made with continuous values for $\beta$, as it becomes apparent form the values given in the respective insets. This leads to an additional deviation between the two, which
is why we display them at equal values for comparison here.

%\newpage

%%%%%%%%% begin colorbox
{ 

%%%%%%%%%%%%%%%%%%%%%%%%%
\section{Effect of a random displacement}\label{Sec:Displacement}

In this appendix we test the sensitivity of our assumption to fit the NN spacings of nests with $\beta$ from a 2D Coulomb distribution to a random displacement of the positions of the nests. 
Because this is an issue in particular for a small number of spacings available, we perform this test for two examples of spacings  of a single year 2020  in Figure \ref{Fig:Displace} for Buzzards (left) and for Goshawks (right). Each year represents an ensemble for each species. We have picked these examples as first, they contain a moderate (228)  respectively small number (19) of spacings, see Fig. \ref{Fig:GroupingYears} top and middle plot of the left column for the undisplaced data. These are reproduced in the top row in Fig. \ref{Fig:Displace}. Second, the two chosen examples give a fitted values of $\beta=1.1$ for the Buzzards and $\beta=3$ for the Goshawks, that is significantly away from the Poisson value of $\beta=0$.
If we were to perform this test on data that are already close to that it would be difficult to see an effect.

The procedure we apply is as follows. First, we determine the mean spacing between the nests of both species separately in 2020, cf. Fig. \ref{Fig:PlotBirds}. We obtain  
$m_B=0.63$ km for Buzzards (dark blue points) and $m_G=2.62$ km for Goshawks (yellow points), before unfolding. 
Next, we pick a random direction for each nest of the ensemble of a given species, say  Buzzard. The nest is then displaced by $Xm_B$ in that direction, where we choose $X=0.5,1.0,1.5$ for that ensemble in three steps. For each chosen value of $X$ we do a fit of $\beta$ to the displaced nests.

The result for the Buzzards in shown Fig. \ref{Fig:Displace} left column with $X$ increasing from top to bottom: the top row corresponds to $X=0$, second row to $X=0.5$, third row to $X=1.0$ and bottom row to $X=1.5$. 
For comparison we give the resulting $\beta$ values (insets) and densities for the Coulomb gas (full line) and for the surmise (dashed line).
The same random displacements and fits for $\beta$ are then made for the ensemble of Goshawks nests, shifted by $Xm_G$ in the rando direction and choosing the same values of $X=0.5,1.0,1.5$, see Fig. \ref{Fig:Displace} right column. As in all previous plots we fit to the cumulative distribution rather than to the density, which would depend on the choice of histogram width.

For the Buzzards, clearly the fitted $\beta$ value rapidly decreases towards Poisson at $\beta=0$. It makes sense that Poisson is only reached once the average mean spacing is exceeded by $X>1$, as only then on average points can become very close.
For the Goshawks, the statistics of 19 spacings in 2020 is (deliberately) poor. Nevertheless, also here the fitted value for $\beta$ consistently moves down towards Poisson, even if the quality of each fit is very low. 

We conclude from this numerical experiment of an increasing random displacement, that our initial data do contain valid information about the repulsion by fitting $\beta$, even down to low statistics with the order of 20 spacings as observed here. For an order of magnitude higher, with about 200 spacings, also the quality of the fits from $\beta$ of order unity down to Poisson becomes acceptable. Poisson is reached when a random displacement larger than the mean spacing is allowed.

\begin{center}
	\begin{figure}[h]
		\centering
		\hspace{0.06\linewidth} \textbf{displaced Buzzard} \hspace{0.1\linewidth}
		\textbf{displaced Goshawks}\\
		\hspace{-4pt}
			\includegraphics[width=0.295\linewidth,angle=0]{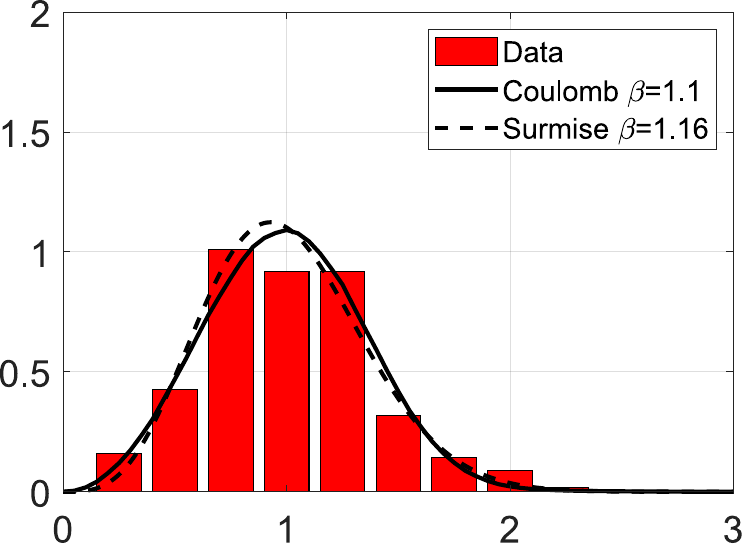}
\hspace{12pt}
			\includegraphics[width=0.295\linewidth,angle=0]{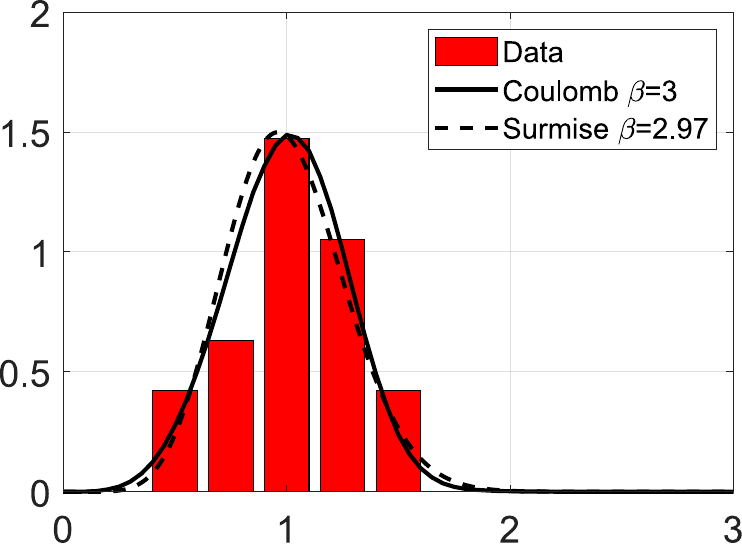}
\\		
		\includegraphics[width=0.3\linewidth,angle=0]{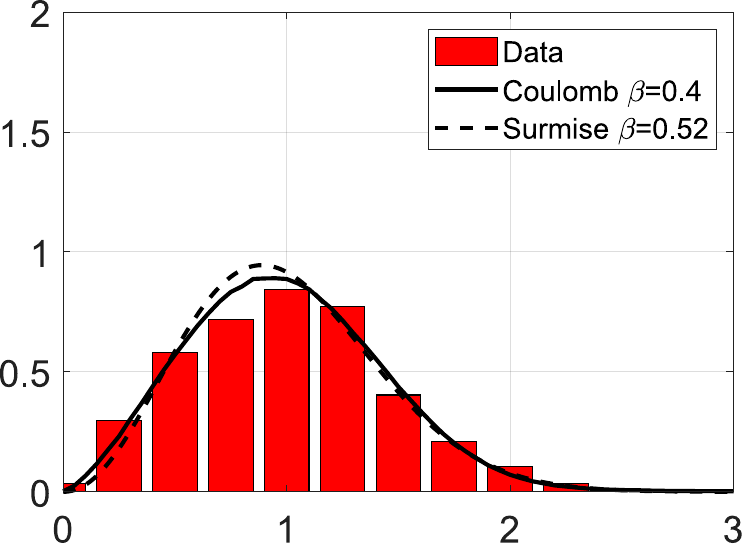}
\hspace{10pt}
		\includegraphics[width=0.3\linewidth,angle=0]{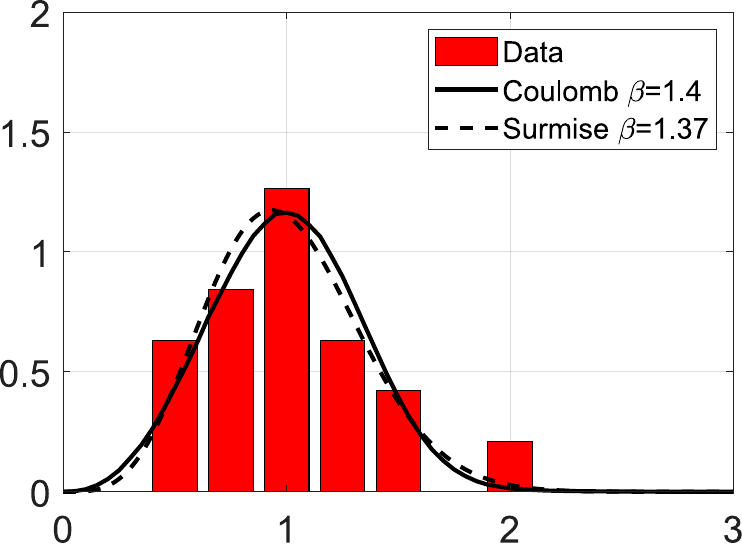}
\\
		\includegraphics[width=0.3\linewidth,angle=0]{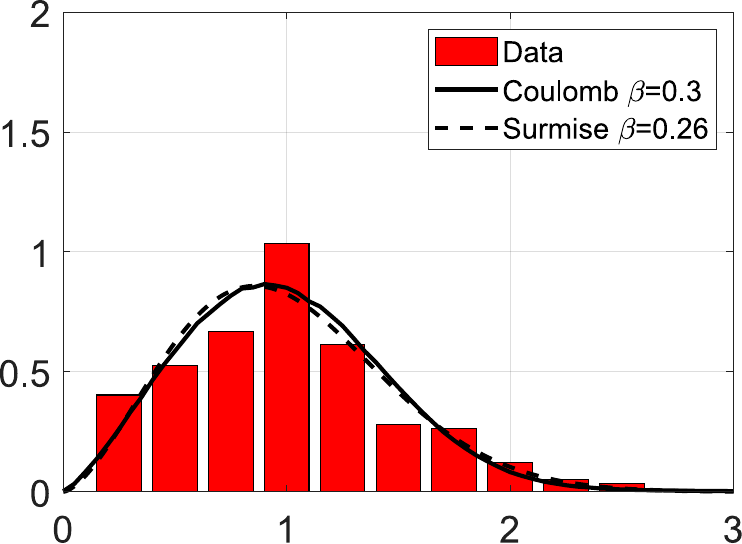}
\hspace{10pt}
		\includegraphics[width=0.3\linewidth,angle=0]{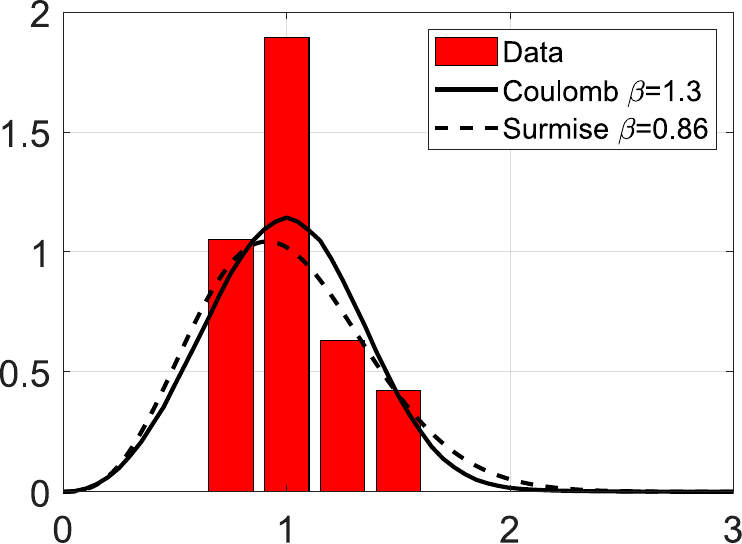}
\\		\includegraphics[width=0.3\linewidth,angle=0]{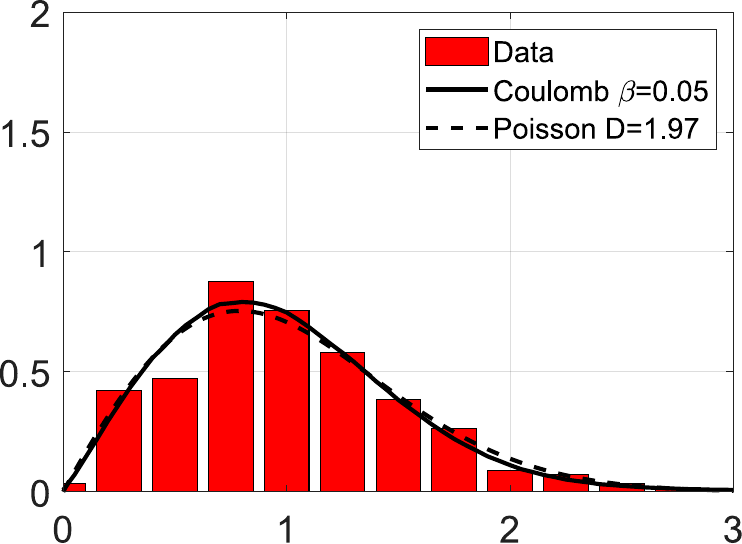}
\hspace{10pt}
		\includegraphics[width=0.3\linewidth,angle=0]{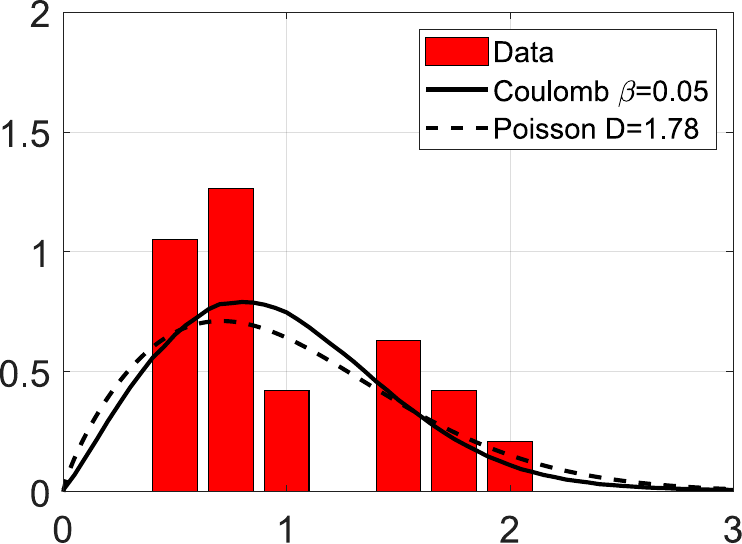}
\caption{
%%%%%%%% 
\textbf{Left row:} Fitted $\beta$ values for NN for randomly displaced Buzzard nests by $Xm_B$: $\beta=1.1\ (X=0)$, $\beta=0.4\ (X=0.5)$, $\beta=0.3\ (X=1.0)$, $\beta=0.05\ (X=1.5)$ from top to bottom. \textbf{Right row:} Fitted $\beta$ values for NN for randomly displaced Goshawk nests by $Xm_G$: 
$\beta=3\ (X=0)$, $\beta=1.4\ (X=0.5)$, $\beta=1.3\ (X=1.0)$, $\beta=0.05\ (X=1.5)$, from top to bottom.
}
		\label{Fig:Displace}
	\end{figure}
\end{center}

\newpage

%%%%%%%%%%%%%%%%%%%%%%%%%%%%%%%%%%%%%%%%%%%%%%%%%%%%%%%
\section{Numbers of nests per species per year}\label{Sec:numbers}

\begin{table}[h!]
%%%%%%%%
	{ 
	\begin{tabular}{c||c|c|c|c|c}
		Year & Buzzards & Goshawks & Eagle Owls & Eagle Owls (F) & Eagle Owls (P)\\
		\hline
		\hline
		2000 & 63 & 12 & 1 & 1 & 0\\
		\hline
		2001 & 85 & 19 & 1 & 1 & 0\\
		\hline
		2002 & 78 & 15 & 2 & 2 & 0\\
		\hline
		2003 & 101 & 17 & 3 & 3 & 0\\
		\hline
		2004 & 99 & 17 & 6 & 6 & 0\\
		\hline
		2005 & 121 & 17 & 7 & 7 & 0\\
		\hline
		2006 & 106 & 18 & 9 & 9 & 0\\
		\hline
		2007 & 107 & 19 & 9 & 9 & 0\\
		\hline
		2008 & 130 & 21 & 7 & 7 & 0\\
		\hline
		2009 & 105 & 20 & 5 & 5& 0\\
		\hline
		2010 & 195 & 22 & 8 & 8 & 0\\
		\hline
		2011 & 161 & 27 & 10 & 8 & 2\\
		\hline
		2012 & 242 & 29 & 12 & 11 & 1\\
		\hline
		2013 & 147 & 22 & 16 & 10 & 6\\
		\hline
		2014 & 220 & 23 & 10 & 10 & 0\\
		\hline
		2015 & 240 & 22 & 15 & 12 & 3\\
		\hline
		2016 & 217 & 22 & 15 & 12 & 3\\
		\hline
		2017 & 267 & 21 & 19 & 13 & 6\\
		\hline
		2018 & 214 & 22 & 16 & 10 & 6\\
		\hline
		2019 & 251 & 19 & 20 & 14 & 6\\
		\hline
		2020 & 228 & 19 & 23 & 16 & 7\\
		\hline
		\hline
		Sum & 3377 & 423 & 214 & 174 & 40\\
	\end{tabular}
	\vspace{10pt}
	\caption{The number of nests per each species per year, including the split of the Eagle owls into (F) and (P) in the last two columns. The sum of all nests per species over all years is given in the last row.}
	\label{tableNumbers}}
%%%%%%%
\end{table}
In Table \ref{tableNumbers} we collect the number of observed nests per species per year that are depicted in Fig. \ref{Fig:AllBirds} bottom left for the Buzzards,  in Fig. \ref{Fig:AllBirds} bottom right for the Goshawks, and in Fig. \ref{Fig:EagleOwlGroup} bottom left respectively right for the Eagle Owls in the forest (F) respectively plain (P). We also give the total number of Eagle Owls per year in the middle column.

%%%%% end colorbox
}

\end{appendix}

%\newpage

%%%%%%%%%%%%%%%

\end{document}